\documentclass[floatfix,onecolumn,nofootinbib,longbibliography,preprintnumbers,11pt]{revtex4}

\usepackage{palatino}

\usepackage{amssymb,amsmath,amsthm}

\newtheorem{remark}{Remark}

\usepackage{natbib}
 
\usepackage{bm} 
\usepackage{graphicx} 
\usepackage{multirow} 
\usepackage{cases} 

\usepackage{booktabs,fancybox}

\usepackage{verbatim}  

\usepackage{xspace}
\makeatletter
\DeclareRobustCommand\onedot{\futurelet\@let@token\@onedot}
\def\@onedot{\ifx\@let@token.\else.\null\fi\xspace}

\def\eg{\emph{e.g}\onedot} 
\def\ie{\emph{i.e}\onedot}

\makeatother

\usepackage[shortlabels,inline]{enumitem}  
\newlist{inlinelist}{enumerate*}{1}  
\setlist[inlinelist]{label=(\roman*)}  
\usepackage{hyperref}

\usepackage{caption}
\usepackage{subcaption}  

\newcommand{\numberfield}[1]{\ensuremath{\mathbb{#1}}} 

\newcommand{\vect}[1] {\ensuremath{\mathbf{#1}}} 

\newcommand{\cvect}[1] {\mathsf{#1}} 

\newcommand{\ev}[1]{\ensuremath{\langle #1 \rangle}} 

\newcommand{\sid}{\ensuremath{_{\text{\tiny\textsf{ID}}}}}   

\newcommand{\psiid}{\ensuremath{\psi\sid}}

\newcommand{\tpsiid}{\ensuremath{\widetilde{\psi}\sid}}

\newcommand{\intrcsonedim}{\ensuremath{\iint_{x_1 \le x_2}}}  

\newcommand{\rcs}{\ensuremath{_\lwr{\text{\tiny\textsf{RCS}}}}}  

\newcommand{\lwr}[1]{{\stackrel{\phantom{.}}{#1}}}

\newcommand{\tableheadtext}[1]{\textsf{\textbf{{\fontsize{8}{9}\selectfont#1}}}} 
\newcommand{\tabletext}[1]{\textrm{{\fontsize{8}{9}\selectfont#1}}} 

\newcommand{\Gmat}[4]{\ensuremath G(#1,#2, #3, #4) }

\begin{document}


\title{Identical Quantum Particles as Potential Parts\footnote{To appear in Phil. Trans. Roy. Soc. (2026), \url{https://doi.org/10.1098/rsta.2025.0360}.}}

\author{Philip Goyal}	
    \email{pgoyal@albany.edu}
    \affiliation{University at Albany~(SUNY), NY, USA}

\date{\today}
\homepage[Homepage:~]{https://www.philipgoyal.org}
\linespread{1.318} 

\begin{abstract}

The mathematical rules used to handle systems of identical quantum particles bring into question whether the elementary constituents of matter, such as electrons, have the fundamental characteristics of persistence and reidentifiability that are usually attributed to classical particles.   However, despite considerable philosophical debate, the metaphysical profile of these entities remains elusive.   

Previous debates have taken the mathematical rules, and the language in which these are usually couched, as a starting point.  Here, we argue that this methodology is inherently limited, and develop a new conception of identical particles based on a recent mathematical reconstruction of these rules.  Using this reconstruction, we demonstrate that the special behaviour of identical particles originates in the confluence of identicality and the active nature of the quantum measurements. We propose that identical particles are appropriately viewed as potential parts of a whole, and show how this leads to striking consequences such as restricted transtemporal identity.

\end{abstract}

\linespread{1.618}

\maketitle

\section{Introduction}

According to the atomistic conception of matter, the physical objects in our everyday environment consist of spatial arrangements of indivisible, indefinitely persistent individuals.  
The atomistic conception runs like a golden thread through the history of physics, being woven into Newtonian mechanics, and inspiring such pivotal discoveries as that of the chemical elements and the subatomic particles of matter.  However, quantum theory has brought the fundamental nature of the denizens of the microscopic realm into question.  For example, the quantum treatment of individual `particles' is widely interpreted to mean that, unlike classical mechanical particles, quantum particles are not continuously localized entities.  Still deeper questions about the metaphysical profile of quantum particles have been raised by the quantum rules for handling systems of identical particles~(i.e. particles with the same state-independent properties)~\cite{French2000}.  In particular, Dirac asserted that, in a system of identical particles, the particles are `absolutely indistinguishable', while Schroedinger claimed that, once part of a system, identical particles suffer a (partial) loss of individuality, in the sense that one cannot speak in an absolute sense of an object observed presently as being `the same' as one observed earlier~\cite[p.~131]{Schroedinger2014}.

The nature of identical particles has been extensively discussed in both the foundations of physics and the philosophy of physics literatures.  However, these discussions have yet to culminate in any consensus in the metaphysical profile of identical particles.  One might suppose that this lack of consensus is due to an unavoidable underdetermination of the metaphysics by the theory.  However, we argue here that there is a more pervasive issue at play, namely that the widely used interpretative methodology introduces ambiguity and empirically unwarranted notions, and pre-emptively excludes certain evidence of potential interpretative import\footnote{A broad discussion of the limitations of the standard interpretative methodology, when applied to quantum theory, is given in~\cite{Goyal2025a, Goyal2025d}.}.  In particular, as we shall show, existing discussions suffer from one or more of the following methodological limitations:
\begin{enumerate}[1.,leftmargin=*]
\item \emph{Discussions of the identical particle formalism in the quantum mechanical setting invariably focus on the symmetrization postulate~(SP) and the symmetric operator constraint~(SOC)}\footnote{The SP imposes a constraint on the allowable states, while the SOC restricts allowable observables to those that are symmetric. Corresponding restrictions of focus occur in philosophical discussions of the Fock space formalism.}.  However, as we shall discuss, the standard mathematical procedure for constructing quantum models of identical particle systems, to which we refer as as the \emph{quantum symmetrization algorithm}~(QSA), is a \emph{multi-step process} that invokes a number of distinct rules in addition to the SP and SOC\footnote{The standard mathematical procedure, as gleaned from the practice of physics, is summarized in Table~\ref{tbl:QSA}.}.  For example, the QSA contains a \emph{no-overlap rule}~(NOR), a heuristic that is used to decide whether or not a quantum model needs to be subjected to symmetrization\footnote{When building models in the context of the Fock space formalism, all states are effectively symmetrized, but a decision is made to drop the SOC where measurements are performed on modes that are deemed orthogonal.}.  
Since empirical test of the SP and SOC require application of the QSA as a whole, any part of the QSA---even a heuristic---could \emph{a priori} be of interpretative relevance.  Nevertheless, discussions assume~(generally implicitly) that components of the QSA other than the SP and SOC are interpretationally irrelevant.

\item \emph{Physicists' reading of the symbolism is generally tacitly accepted, without establishing whether or not that reading is empirically justified.}  For example, the indices in symmetrized states such as~$\psi(x_1, x_2) = [\alpha(x_1)\beta(x_2) \pm \beta(x_1)\alpha(x_2)]/\sqrt{2}$ are routinely read as \emph{particle} labels.  But, as we shall discuss~(Sec.~\ref{sec:MPI}), this reading is questionable given the rules' context of successful application.  

\item \emph{Discussions of the mathematical rules for handling systems of identical particles generally do not clearly distinguish between the rules themselves, on the one hand; and theoretical justifications, or philosophical interpretations, of these rules, on the other.}  For example, the common statement that `identical particles are indistinguishable' is generally taken to be a factual statement about the identical particle quantum formalism on a par with the SP itself%
\footnote{For example: \begin{inlinelist} \item ``It is a fundamental principle of quantum mechanics that two particles of the same kind are absolutely indistinguishable.''~\cite{ChiaraFrancia1992};  \item ``If there is any consensus as to what particle indistinguishability means, it is its formal expression in quantum mechanics: particles must have exactly the same mass, spin, and charge, and their states must be symmetrised, yielding either symmetric or antisymmetric wave-functions.''~\cite[p.1]{Saunders2020}. \end{inlinelist}}.
However, as we shall see, this statement is not part of the QSA, and is instead properly viewed as a partial justification, interpretation, and consequence~(modulo additional interpretative assumptions) of parts of the symbolism~(Sec.~\ref{sec:MPI} and~\ref{sec:MI}).  It particular, the statement strictly speaking goes beyond what is necessary to build empirically successful models of identical particle systems, and accordingly has no empirical warrant \emph{per se}.

\item \emph{Assumptions implicit in experimental study of identical particles are not subjected to systematic analysis and interpretation on a par with the formalism.}  For example, in a bubble chamber image, the bubbles are parsed into \emph{tracks}, and the geometry of these tracks is then further processed to yield information about the state-independent properties~(such as mass and charge) of the particles assumed to generate these tracks.  Implicit in this parsing is the assumption that, in this experimental context, these tracks are generated by \emph{persistent, reidentifiable individuals}.  However, such assumptions are rarely given any interpretational consideration\footnote{For example, Chiara and di Francia dismiss the operational assumption of persistence as `mock persistence' on the grounds that it exists for a `very brief duration'~\cite[p.~266]{ChiaraFrancia1992}.}.
\end{enumerate}




Some of the above-mentioned shortcomings have been recognized in the literature.  For example, even while asserting~(on the basis of Bose's algorithm for deriving quantum statistics) that identical particles are non-individuals~(\emph{`the elementary particle is not an individual; it cannot be identified, it lacks `sameness.'}~\cite[p.~109]{Schroedinger1950}), Schroedinger recognizes that a `restricted notion of identity' is needed to make sense of particle tracks~\cite[p.~114-115]{Schroedinger1950} and introduces a precursor of the no-overlap condition~\cite[\S13]{Schroedinger1950}.  This idea has been recently echoed by Bigaj~\cite{Bigaj2020b}, who argues that, in certain circumstances, identical particles can be physically labelled~(and thus reidentified), and proposes a notion of weak transtemporal identity~\cite[\S5]{Bigaj2020b} or partial genidentity~\cite[Ch.~6]{Bigaj2022}.   A related thread has been taken up by Dieks and Lubberdink~\cite{DieksLubberdink2011,DieksLubberdink2020,Dieks2020}, who posit the emergence of classical particle tracks in the classical limit of a system of identical quantum particles, and suggest that this emergence requires that one drop the heretofore near-universal assumption that the indices in symmetrized states refer to individual particles, a suggestion that has received significant recent attention~\cite{Caulton2014,MullerLeegwater2020,Friebe2022}.

However, as valuable as these insights are, the underlying interpretative methodology remains unchanged, \emph{viz.} a subset of the mathematical procedures for modelling systems of identical particles~(whether that be in the first- or second-quantized formalism), together with most of the language in which these procedures are customarily couched, are taken as the starting point for philosophical reflection.  As described above, this interpretative approach is inherently problematic:
\begin{enumerate}
\item Insofar as one deems the identical particles formalism worthy of interpretative \emph{because} of its empirical success, \emph{a priori} no part of the QSA can be assumed to lack interpretational significance.
\item The QSA is described using a mixture of mathematical and natural language.  The natural language consists of description of the symbolism~(such as indices~$i$ of~$\psi(x_1, x_2)$ described as `particle labels') together with other descriptive language~(such as `particle' and `system').  Presentations of the QSA are also invariably accompanied by additional language~(such as the claim that identical particles are `indistinguishable').  Presumably some of this language is necessary to apply the QSA to construct quantum models, while the remainder is not.  However, as we shall see~(Sec.~\ref{sec:QSA}), it is exceedingly difficult to establish precisely what parts of this natural language has genuine empirical warrant.

\item Empirical procedures are essential to connect a mathematical model to the sensory realm, and thereby establish empirical validity  of the QSA.  Since the quantum realm involves objects which are far removed from those encountered in classical physics~(or indeed everyday life), these procedures are sophisticated, and involve non-trivial assumptions in both their design and in the interpretation of the data they generate.  Hence, \emph{a priori}, these procedures, and the assumptions implicit therein, cannot be assumed to be interpretationally insignificant. 
\end{enumerate}

To avoid these problems, we maintain that an interpretative methodology is required which does not take the existing identical particle formalism as a starting point.  Accordingly, here we adopt a new interpretative methodology~\cite{Goyal2025a} based on a \emph{mathematical reconstruction}~\cite{Goyal2015} of the Feynman rules for handling identical particles~(hereafter \emph{Feynman quantum symmetrization algorithm}, or fQSA).   Broadly, reconstruction of the quantum formalism~(or a part thereof) involves the formulation of physically well-motivated assumptions and postulates expressed in an operational framework, and in the mathematical derivation of the formalism from these assumptions and postulates~\cite{Grinbaum-Reconstruction-Interpretation, Grinbaum-Reconstruction, Hardy2013,Goyal2022c}.  The reconstruction program has had substantial success over the last 25 years~\cite{Hardy2013,Goyal2022c}, and has attracted growing interest from philosophers~\cite{Grinbaum-Reconstruction-Interpretation,Grinbaum-Reconstruction,Stairs2015,Dickson2015,Felline2016,BerghoferGoyalWiltsche2019,Berghofer2023,Oddan2024,Berghofer2024}.   
As described in~\cite{Goyal2022c,Goyal2025a,Goyal2025d}, the success of the reconstruction program opens up a \emph{reconstruction-based interpretative methodology}:~with a reconstruction in hand, interpretation proceeds by reflecting on the operational framework, and on the additional assumptions and postulates that are employed to derive the formalism.   The methodology of reconstruction is of particular interpretative value since it distills the full physical content of an empirically validated formalism into a set of assumptions and postulates couched in language that is more amenable to philosophical reflection, and its operational framework articulates the minimum experimental capabilities that are necessary to connect the formalism to the sensory realm.

In the present instance, the reconstruction of the fQSA substantially ameliorates each of the three above-mentioned difficulties arising from taking the QSA as a given.  First, the Feynman quantum symmetrization procedure is reconstructed on the basis of a \emph{single} core postulate, the so-called operational indistinguishability postulate~(OIP), supported by a single operationally well-motivated auxiliary condition~(the so-called \emph{isolation condition}).  Hence, the OIP becomes the primary target for interpretation, instead of the amalgam of separate rules that comprise the QSA.  Second, the reconstruction takes place in an operational framework, and the OIP and auxiliary assumptions are expressed in terms of operational notions such as momentary, localized detection events. 
One thereby minimizes the risk of overlooking implicit interpretative assumptions, or of conflating non-trivial interpretative claims with the minimally-necessary interpretative assumptions.  Third, as the fQSA is derived in direct contact with basic experimental operations, it is much easier to unearth---and take fully into account---key assumptions implicit in experimental practice.

The above-mentioned mathematical reconstruction of the fQSA has recently attracted significant interpretational interest~\cite{Goyal2019a,Jantzen2020,Bitbol2023}.  Here we build upon Ref.~\cite{Goyal2019a} to develop a detailed metaphysical interpretation.  The interpretation is developed in three main steps.  

First, we carry out an operational analysis of the relationship between reidentification and persistence in classical and quantum systems.  Our key finding is that, in a quantum setting, a system of identical particles suffers an \emph{in-principle loss of reidentifiability} due to the \emph{confluence of particle identicality and the active nature of quantum measurement}, which removes direct empirical justification for the metaphysical assumption that two particles continue to individually persist \emph{during} collision.  

Second, we re-interpret the reconstruction described in~\cite{Goyal2015}~(and partially interpreted in~\cite{Goyal2019a}) as a formal means of \emph{relaxing} the notion of persistent particle without eliminating it entirely.   In the reconstruction, this is achieved by taking \emph{detection-events}~(rather than particles) as the primary reality, and then formulating two distinct and complementary object-models---the so-called \emph{persistence} and~\emph{nonpersistence} models~\cite{Goyal2019a}---of the same experimental data.  These models differ in the persistent object(s) which they posit as underpinning the detection events:~the persistence model posits that the detection-events are underpinned by persistent particles, while the nonpersistence model posits that the system as a whole---a single holistic object~(an extended simple)---is responsible.

Third, in Sec.~\ref{sec:potential-parts}, we argue that the complementarity object-models are most appropriately understood via the metaphysical doctrine of \emph{potential parts}, which leads to the following proposal.  Identical particles do not simply \emph{exist} in a context-independent sense, as per the atomistic conception of matter.  Rather, when sufficiently isolated~(in, say, separate particle traps), identical particles \emph{can} be said to exist as actual parts of a system.  If, however, one considers two electrons in a helium atom, whose behaviour cannot be understood using the persistence model alone, the electrons exist \emph{potentially}.  Furthermore, there exists a physical process of division~(complete ionization) which gives rise to two (actually existing)~electrons. In addition, the helium system manifests a holistic object~(described by the nonpersistence model), which ceases to exist upon complete ionization.  As we discuss, this viewpoint has some striking consequences, including weak transtemporal identity.

The paper is organized as follows.  In Sec.~\ref{sec:QSA-and-difficulties}, we give an explicit statement of the QSA as gleaned from the practice of physics, discuss its minimal physical interpretation, and analyse the difficulties encountered in its metaphysical interpretation.  In Sec.~\ref{sec:persistence-and-reidentification}, we carry out an operational analysis of persistence and reidentifiability in classical and quantum systems.  In Sec.~\ref{sec:persistence-nonpersistence}, the reconstruction of the fQSA is summarized and analysed, and in Sec.~\ref{sec:potential-parts}, the connection to the metaphysical doctrine of potential parts is explicated.  We conclude in Sec.~\ref{sec:conclusion} with a summary and some open questions.

\section{The quantum symmetrization algorithm:~incompleteness and interpretational difficulties}
\label{sec:QSA-and-difficulties}

The general mathematical rules for handling systems of identical particles were assembled in three main stages.  First, in 1926, Dirac~\cite{Dirac1926} and Heisenberg~\cite{Heisenberg26} showed that the previously established statistics of Bose-Einstein~\cite{Bose1924,Einstein1925} and Fermi~\cite{Fermi1926}~(the latter based on Pauli's exclusion principle~\cite{Pauli1925}) could be incorporated within the nascent quantum formalism by restricting states to those that are symmetric or antisymmetric~(hereafter the \emph{symmetrization postulate}~(SP)).  Second, two auxiliary conditions were formulated---(i) a restriction of physically meaningful observables to those that are symmetric~(SOC)~\cite{Dirac30}; and (ii)~a criterion to determine whether a system requires that its state space and observables be restricted, \emph{viz.} the aforementioned no-overlap rule~(NOR)~(e.g.~\cite{Schroedinger1950}).  Third, the spin-statistics theorem~\cite{Fierz1939,Pauli1940} established the existence of a fundamental relation between the intrinsic properties of the identical particles in a system and the symmetry type of its states.  

These rules form the basis of a \emph{de facto} algorithm for constructing a quantum model of a system of identical quantum particles, to which we refer~(with the exclusion of the spin-statistics theorem) as the \emph{quantum symmetrization algorithm}~(QSA).  To the best of our knowledge, this algorithm has not previously been stated in its entirety.  Physics monographs and quantum theory textbooks~(at all levels) generally articulate the SP and to a lesser extent the SOC.  However, presentation of the other parts of the QSA is uneven.  For example, although any application of the SP requires that one make a non-trivial decision on what constitutes the `system' of interest,  the system-demarcation criteria are usually conveyed to the reader \emph{implicitly} through example problems.  In those few instances where the NOR is stated outright, its \emph{status} vis-\`a-vis the SP is ambiguous:~in some cases, it is presented as a stand-alone criterion that in necessary to decide whether or not to symmetrize a quantum state~(e.g.~\cite{Schroedinger1950}); in others, it is presented as if it were derivable as a \emph{special case} of the SP~(\eg~\cite[Ch~14, \S8]{Messiah1962V2} and~\cite[p. 1406--1408]{CohenTannoudjiV2}).  However, since such derivations involve a number of non-trivial interpretative assumptions~(such as indices interpreted as particle labels), their validity is far from clear.

Here, we do not attempt to \emph{resolve} the ambiguities of what precisely constitutes the QSA or what parts of it can be derived from other parts.   Rather, our aim is to present the procedure that is routinely followed~(even though not all its steps are commonly articulated) in the construction of quantum models of identical particle systems.  This suffices for our present purposes in this Section, which is to lay out the difficulties that are faced when attempting to interpret the QSA, and to show that it is \emph{not a reliable basis for establishing the nature of identical quantum particles}\footnote{ As explained in the Introduction, it is precisely due to the ambiguities of the QSA that our interpretation of the nature of identical particles~(developed in subsequent sections) will be based on an operational reconstruction of the fQSA.  The reconstructed fQSP can be reformulated as a state-based quantum symmetrization procedure~(QSP), which can replaces the QSA~\cite{Goyal2015,Goyal2019a}.  A detailed discussion of the QSP, and its comparison to the QSA, is beyond the scope of this paper.  However, for completeness, the QSP arising from the reconstructed fQSP is briefly outlined in Appendix~A.}. For a similar reason, we do not discuss the quantization algorithm when expressed in the Fock space formalism---this algorithm, although subtly different from the QSA, suffers analogous ambiguities.

\subsection{Distinguishing the QSA from justificatory and interpretative statements}

Since its inception, physicists' descriptions of the rules that comprise the QSA have been invariably accompanied by: 
\begin{enumerate} 
	\item \emph{Justificatory statements} that claim or imply that one or more of these rules can be \emph{accounted for} or \emph{derived} on the basis of some key idea;

	\item \emph{Physical interpretative statements} that ascribe physical meaning to parts of the symbolism; and 
	
	\item \emph{Metaphysical interpretative statements} that make claims as to the nature of identical particles
\end{enumerate}
For example, Dirac begins his systematic development of the formalism for handling systems of identical particles~\cite[\S62]{Dirac30} by asserting that identical particles in a system are `absolutely indistinguishable', and implies that the curious behaviour of identical quantum particles can be directly traced back to this idea:
\begin{quote}
``If a system in atomic physics contains a number of particles of the same kind, e.g. a number of electrons, the particles are absolutely indistinguishable one from another.  No observable change is made when two of them are interchanged. This circumstance gives rise to some curious phenomena in quantum mechanics having no analogue in the classical theory [...]'''~\cite[\S62]{Dirac30}.  
\end{quote}
In this instance, `particle indistinguishability' plays the dual role of justification and metaphysical interpretation.  With respect to metaphysical interpretation, Dirac's notion of \emph{interchanging} implies that he views the particles as individuals~(with trans-temporal identity), while the assertion of \emph{no observable change} implies the impossibility of reidentification\footnote{The notion of reidentification is discussed in Sec.~\ref{sec:persistence-and-reidentification}.} of any such interchange.

This tendency persists to the present day, amongst both physicists and philosophers of physics, in such claims as `identical particles are indistinguishable'.  However, such claims are unnecessary for the application of the mathematical rules themselves---as we shall see below, one can state the QSA without reference to indistinguishability.   We also remark that the notion of `indistinguishability' is itself interpreted or formalised in a variety of distinct ways by different authors\footnote{For example, `indistinguishable' is sometimes taken to mean:~\begin{inlinelist} \item two particles possess the same intrinsic properties~(\ie it is deemed synonymous with physicists' usage of `identical'); \item two particles possess the same intrinsic \emph{and} state-dependent properties, the latter typically formalized by the permutation-invariance condition~(see Remark~\ref{rem:justification-of-SOC}(b)); \item a physical system returns to the same state of affairs~(or the same mathematical state) after a `swap' of two identical particles;  \item after a `swap', a physical system has a state~(or alternately a state of affairs) which cannot be distinguished by any observer from the prior state (of affairs), even though the prior and posterior states~(of affairs) differ;  \item an observer is unable to reidentify two identical particles due to his (possibly in-principle) inability to observe a continuous `swap' of the particles; \item a system of identical quantum particles is in~(or must be in) a symmetric or antisymmetric quantum state~(\ie it is deemed synonymous with the SP). \end{inlinelist}  As indicated above, Dirac's own words suggest that by `absolutely indistinguishable', Dirac meant that \begin{inlinelist} \item identical particles are persistent individuals; yet \item under no circumstances can they be reidentified by any observer.  See~\cite{Saunders2020} for a recent attempt to establish Dirac's intended meaning of the concept. \end{inlinelist}}, which complicates any discussions that invoke this notion.  In what follows, we shall endeavour to indicate which meaning is intended wherever relevant~(in particular, see Remark~\ref{rem:justification-of-SOC}).

In order to place our discussion of the quantum rules for handling identical particles on a secure footing, we strive to distinguish between:
\begin{enumerate}
\item The quantum symmetrization algorithm~(QSA), namely the mathematical recipe that is followed when building quantum models of identical particle systems.
\item The minimal physical interpretation~(MPI) of the algorithm, which provides the \emph{minimum physical meaning} to the symbolism that one needs in order to build theoretical models whose predictions can be compared to experiment.
\item A metaphysical interpretation of the algorithm~(MI), which proposes an ontology~(however partial) of the physical reality described by the QSA.
\end{enumerate}
The QSA and MPI together provide the physicist with the guidance necessary to build empirically-testable models.  As the QSA and MPI have been battle-tested in experiment, they have strong empirical warrant. Any metaphysical interpretation~(MI) is, on the other hand, relatively underdetermined by experiment, and is therefore more speculative.  However, as we shall see below, it is far from straightforward to determine---by inspection of the given formalism and presentations thereof---which physical assertions belongs to the MPI.  Hence, the main conclusion of this section is that any attempt to establish the metaphysical profile of identical quantum particles on the basis of interpretation of the QSA~(or any variant thereof) is inherently problematic.

Below, we outline the QSA~(Sec.~\ref{sec:QSA}) and discuss the MPI~(Sec.~\ref{sec:MPI}).  We then briefly discuss some of the interpretations---both common and recently advanced---and their weaknesses~(Sec.~\ref{sec:MI}).

\subsection{The quantum symmetrization algorithm}
\label{sec:QSA}

The construction of a quantum model of a system of $n$ identical particles takes place in three steps~(summarized in Table~\ref{tbl:QSA}, with example in Table~\ref{tbl:QSA-examples}):
%
\linespread{1.218} 
\setlength{\tabcolsep}{3pt}
\begingroup
\begin{table*}
\centering
	\begin{ruledtabular}
    \begin{tabular}{p{3.5cm}p{5.25cm}p{5.50cm}} 
    \smallskip
    {\bf }  															& \tableheadtext{Description} 			 & \tableheadtext{Remarks}    \\ \hline

\smallskip\tableheadtext{I. Foil model~(FOIL)}
																	&	\smallskip\tabletext{Construct quantum model of system, 
																									  without according the identical particles any special treatment.}   	
																	&	\smallskip\tabletext{1. At this stage, `system' is user-defined, 
																					   without explicit criterion on how the choice is made.}
																		\newline 
																		\tabletext{2. The classical Hamiltonian is symmetric in particle labels 
																						since the particles are identical.}
																		\newline 
																		\tabletext{3. A particle-label reading of indices~(PLI) and a persistent particle
																						ontology~(PPO) are inherited from the classical model of the system.}
																	\\
    
\smallskip\tableheadtext{II. No-overlap rule~(NOR)}  
																	&	\smallskip\tabletext{If the particles' wavefunctions~(or reduced states) do not overlap, 
																									  the foil model suffices.}
																	&	\smallskip\tabletext{1. Reduced states must be considered if the system state is entangled.}
																		\newline
																		\tabletext{2. If the foil model suffices, no constraints on states or observables are 
																									  imposed---all states are valid, and observables such as~$\hat{x}_1$ are permitted.}
																	\\   

\smallskip\tableheadtext{III. Symmetrization~(SYM)}   						
																	&	\smallskip\tabletext{\emph{If NOR is not satisfied:}} \newline
																		\tableheadtext{1. Symmetrization postulate~(SP).} 
																		\newline
																		\tabletext{(Anti-)symmetrize the system wavefunction.}
																		\newline 
																		\tableheadtext{2. Symmetric Observable Constraint~(SOC).}
																		\newline
																		\tabletext{Restrict observables to those that are symmetric.}
																	&	\smallskip\tabletext{1. Whether to symmetrize or antisymmetrize is determined by 
																									  the type of identical particle via the spin statistics theorem~{(SS)}.}
																		\newline
																		\smallskip\tabletext{2. The necessity~(and rationale) for the SOC is disputed.}
																		 
																	\\

	\\
    \end{tabular}
	\end{ruledtabular}
\caption{\label{tbl:QSA} \emph{Quantum Symmetrization Algorithm~(QSA).}}
\end{table*}
\endgroup
\linespread{1.618} 
\begin{enumerate}[I.,leftmargin=*]
\item\textbf{Construct Foil Model~(FOIL).}
Construct a provisional---or \emph{foil}---quantum model of the system, taking no special account of the fact that its constituent sub-systems~(`particles') are identical.  Construction of this foil model follows standard procedures for constructing quantum models of physical systems.  For example, one typically obtains a foil model of the electrons in an atom by writing down the \emph{classical} model of the system~(in particular its classical state and Hamiltonian, including spin terms) and then judiciously applying classical-quantum correspondence heuristics to obtain the foil quantum model.  

\begin{remark} The particles, being identical, have the same state-independent properties, so that the classical Hamiltonian,~$H(x_1, p_1; x_2, p_2; \dots; x_n, p_n)$, is a symmetric function of the particles' state-dependent properties; \emph{viz.} for any permutation~$\pi$,
\begin{equation}
H(x_{\pi(1)}, p_{\pi(1)}; \dots; x_{\pi(n)}, p_{\pi(n)}) =  H(x_1, p_1; \dots; x_n, p_n).
\end{equation}
The quantum Hamiltonian,~$\hat{H}(\hat{x}_1, \hat{p}_1; \dots; \hat{x}_n, \hat{p}_n)$, is correspondingly symmetric.  Hence, in the foil model, particle identicality is reflected solely in the symmetry of the quantum Hamiltonian---there is as yet no restriction on states or measurement operators. 
\end{remark}

\begin{remark}
In the foil model, the operators~$\hat{x}_i$ and~$\hat{p}_i$ are interpreted as associated with particle~$i$, to which we shall refer as a \emph{particle-label reading of indices}~(PLI).  In addition, the foil model inherits an ontology---the \emph{persistent particle ontology~(PPO)}---from the classical model, so that index~$i$ refers to a \emph{persistent individual}, namely~\emph{particle}~$i$.
\end{remark}

\item\textbf{Apply No-Overlap Rule~(NOR).}
The \emph{no-overlap rule}~(NOR) posits that the foil model of a system suffices provided that the wavefunctions~(or reduced states) of the particles do not overlap.  For example, two electrons in separate hydrogen atoms at ordinary room conditions satisfy NOR to a very high approximation, as do two entangled photons propagating in separate modes in a Bell experiment.  However, under the same conditions, two electrons in a helium atom fail to satisfy NOR, so that the foil model is inadequate.   

\begin{remark}
Dirac's original formulation~(`if a \emph{system} contains many particles of the same type....'~\cite[\S62]{Dirac30}), presumes that the user has already sensibly analysed the physical situation of interest into `systems'.  However, as Schroedinger~\cite{Schroedinger1950} and others later recognized, making sense of situations such as particle tracks in a bubble chamber require a system-demarcation criterion since different analyses will, in general, yield different results.  NOR provides such a criterion.  For example, in a box of hydrogen gas at ordinary conditions, each hydrogen atom can be regarded as a `system', for which the foil model suffices.  Hence, the joint system comprising these two atoms can be obtained by taking tensor products of their respective foil models.  However, each electron in a helium atom cannot, according to NOR, be regarded as a `system' correctly describable by a foil model.  Hence, the `system' boundary must be extended to include both electrons.   
\end{remark}

\begin{remark}
As indicated, the NOR is an approximate rule.  In practice, in the foil model, the reduced states of two electrons associated with different hydrogen atoms will overlap to some degree irrespective of their separation.  For this reason, one might be inclined to drop NOR entirely and proceed with Step~III.  However, if NOR were dropped entirely, then SP and SOC together imply that there is no measurement operator which is capable of describing a position measurement of a particle in a given hydrogen atom.  As indicated in Remark~\ref{rem:justification-of-SOC}~(below), SOC also apparently cannot be dropped without unphysical consequences.
\end{remark}

\item\textbf{Apply Symmetrization to States and Observables~(SYM).}
If NOR is not satisfied, the symmetrization step imposes restrictions on states and measurement operators:
	\begin{enumerate}
	\item\emph{Symmetrization Postulate~(SP).} Restrict consideration to states that are \emph{symmetric} or \emph{antisymmetric}, with the choice of symmetry to be determined by the type of particle under consideration; and 
	\item\emph{Symmetric Operator Constraint~(SOC).} Restrict consideration to measurement operators~(observables) that are \emph{symmetric}.
	\end{enumerate}

\begin{remark}
As mentioned in Step~I, the quantum Hamiltonian of the foil model is symmetric.  However, the SOC asserts that \emph{every} valid observable must be symmetric, which rules out nonsymmetric observables that are admissible in the foil model, such as~$\hat{x}_1$ or~$(\hat{x}_2 - \hat{x}_1)$.  \end{remark}

\begin{remark}\label{rem:justification-of-SOC}
The rationales for the SOC found in the literature all implicitly depend upon the aforementioned interpretation of indices as particle labels, which here manifests in the interpretation of measurement operators---\eg the operator~$\hat{x}_i$ is taken to represent a measurement upon particle~$i$.  Most authors insist upon the SOC, but the commonly-provided rationales are problematic.  For example:~
\begin{enumerate}
\item\emph{Prevention of unphysical expectation values.}  One rationale is that one must exclude non-symmetric operators such as~$\hat{x}_1$, since they yield unphysical expectation values on symmetrized states\footnote{For example, the expectation value of~$\hat{x}_i$ in any (anti-)symmetric state is independent of~$i$.  On the particle-label reading of~$i$, this conflicts with the intuition that the expected locations of the particles will, in general, differ.}.  However, such operators are both meaningful and necessary when NOR does apply~(for example, given two positrons in two separate infinite potential wells, the operator~$\hat{x}_1$ refers to a position measurement on positron~$1$).  And since NOR never applies exactly in any real-world situation~(for example, to two positrons in actual traps), it is difficult to see why~$\hat{x}_1$ abruptly ceases to be valid whenever NOR strictly fails to hold.  

\item\emph{SOC from permutation invariance.}  \label{rem:item:permutation-invariance} A second rationale is that SOC follows from `indistinguishability', formalized as the requirement that a physically-meaningful observable,~$\hat{A}$, yield the same expected value on any permutation of any state,~$\ev{\hat{A}}_\psi = \ev{\hat{A}}_{\hat{P}\psi}$ for all permutation operators~$\hat{P}$ and all states~$\psi$, where `all states' is taken to include both symmetric and non-symmetric states\footnote{In~\cite[\S64]{Dirac30}, Dirac considers permutations as observables in the context where no state restrictions are imposed.  This lack of state restriction is adopted by Messiah and Goldberg~\cite{MG64} in their formalization of `indistinguishability' as they seek to show that this notion does not lead to the SP.}.  However, if one has already granted the SP, the requirement `all states' is questionable.  If the indistinguishability condition is instead restricted to `all (anti-)symmetric states', then it is satisfied by \emph{every} operator~(not just symmetric ones), but one is then faced with the above-mentioned difficulty of physically interpreting non-symmetric operators.      
\item\emph{Prevention of non-symmetric post-measurement states.}  A third rationale arises from the observation that if it were possible to perform a measurement of a non-symmetric observable, such as~$x_1$, then one could put the system in non-symmetric state, violating SP.  However, it is unclear why one could not stipulate that the post-measurement state be (anti-)symmetrized to conform with the SP.
\item\emph{Preclusion of unphysical measurements due to lack of reidentifiability.}  A fourth rationale is based on the idea that  `indistinguishability' implies that no measurement can be performed which probes the properties of a given particle, since such a measurement would require that we identify a particle based only on its label.  However, this presumes that one cannot reidentify a particle even probabilistically on the basis of its state-dependent properties.  Since such reidentification is routinely presupposed in analysis of data~(\eg in the parsing of a bubble chamber image into `particle tracks'), the assertion that no reidentification whatsoever is possible is questionable.
\end{enumerate}
\end{remark}

\end{enumerate}

\linespread{1.218} 
\setlength{\tabcolsep}{10pt}
\begingroup
\begin{table*}
\centering
	\begin{ruledtabular}
    \begin{tabular}{p{2.55cm}p{6.00cm}p{6.00cm}} 
    \smallskip
    {\bf }  															& \tableheadtext{General case} 			 & \tableheadtext{Two structureless identical particles}    \\ \hline

\smallskip\tableheadtext{I. Foil model}
																	&	\smallskip\tabletext{$H_{\text{clas.}} \rightarrow \hat{H}$, which acts on~$\mathcal{H} \equiv \mathcal{H}_1 \otimes\dots\otimes\mathcal{H}_n.$}
																		\newline																		
																		\smallskip\tabletext{$\hat{H}\cvect{v} = i\,\hbar\,\dot{\cvect{v}}\rightarrow \text{Solve for~$\cvect{v}.$}$ }      	
																		&	\smallskip\tabletext{$H_{\text{clas.}}(x_1, p_1; x_2, p_2) = \frac{p_1^2}{2m} + \frac{p_2^2}{2m} + V(x_1, x_2)$}
																			\newline
																			\smallskip\tabletext{$\rightarrow\hat{H}(\hat{x}_1, \hat{p}_1; \hat{x}_2, \hat{p}_2) = \frac{\hat{p}_1^2}{2m} + \frac{\hat{p}_2^2}{2m} + V(\hat{x}_1, \hat{x}_2)$.}
																			\newline
																			\tabletext{Solve~$\hat{H}\psi = i\hbar\partial\psi/\partial t$ for given~$V$.}
																			\newline
																			\tabletext{\emph{(a) Particles in separate traps:}}
																			\newline
																			\tabletext{$\psi(x_1, x_2) = \sum_{ij} c_{ij}\phi_i(x_1)\,\chi_j(x_2)$,}
																			\newline
																			\tabletext{where~$\phi_i$ and~$\chi_j$ are sets of energy eigenstates for each trap.}
																			\newline
																			\tabletext{\emph{(b) Two particles in one trap:}}
																			\newline
																			\tabletext{$\psi(x_1, x_2) = \sum_{ij} c_{ij}\omega_i(x_1)\,\omega_j(x_2)$,}
																			\newline
																			\tabletext{where~$\omega_i$ is a set of one-particle energy eigenstates for trap.}

																	\\
    
\smallskip\tableheadtext{II. No-overlap rule}  
																	&	\smallskip\tabletext{If the reduced states~$\rho_i, \rho_j$ do not overlap for all}
																		\newline\tabletext{$i \neq j$ for given~$\cvect{v}$, then the foil model suffices.}
																	&	\smallskip\tabletext{(a) Reduced states~$\rho_1, \rho_2$ are nonoverlapping for any~$\psi(x_1, x_2)$, so foil model suffices.}
																		\newline
																		\tabletext{(b) Reduced states typically overlap~(or rapidly evolve to overlap), so Step III needed.}
																	\\   

\smallskip\tableheadtext{III. Symmetrization}   						
																	&	\smallskip

																		\tabletext{1. Replace~$\cvect{v} \in \mathcal{H}$ with corresponding~$\cvect{v}_{\pm} \in \mathcal{H}_{\pm}$, where~$\mathcal{H}_{\pm}$ is the (anti-)symmetric subspace of~$\mathcal{H}$.}
																		\newline 

																		\tabletext{2. Restrict observables to~$\hat{A}$ satisfying~$\ev{\hat{A}}_\cvect{v} = \ev{\hat{A}}_{\hat{P}\cvect{v}}$ for all permutation operators~$\hat{P}$.}
																	&	\smallskip
																		\tabletext{1. Replace~$\psi(x_1, x_2)$ with $\text{Proj}_\pm\left(\psi(x_1, x_2)\right)$, where~$\text{Proj}_\pm(\cdot)$ is the normalized projection onto the (anti-)symmetric subspace.} 
																		\newline
																		\smallskip\tabletext{2. Restrict observables to symmetric functions of~$\hat{x}_1, \hat{x}_2, \hat{p}_1, \hat{p}_2$.}
																		 
																	\\

	\\
    \end{tabular}
	\end{ruledtabular}
\caption{\label{tbl:QSA-examples} \emph{Elementary Application of the Quantum Symmetrization Algorithm.}  
}
\end{table*}
\endgroup
\linespread{1.618}

\subsection{Minimal Physical interpretation of the QSA}
\label{sec:MPI}

The above statement of the QSA makes no use of the notion `indistinguishable', yet suffices for the construction of physical models of systems of identical particles such as atoms, molecules, and materials.  Hence, the common claim that `identical particles are indistinguishable' belongs not to the MPI but to an attempt to justify part of the QSA~(see Remark~\ref{rem:justification-of-SOC}) or to a MI~(see Sec.~\ref{sec:MI}).

However, the statement of the QSA \emph{does} make use of a number of idealized theoretical terms---`system', `particle', and so forth.  These terms guide the application of the QSA in non-trivial ways, and influence its metaphysical interpretation.  Hence, it is important to analyse their meaning in order to establish what aspects of their meaning is warranted by the empirical success of the QSA.   To this end, we can divide the \emph{physical meaning} of each term into an \emph{operational meaning} and a \emph{metaphysical meaning}.  In the context of the QSA, the physical meaning of  `particle' and `system' are of particular interest.  

\subsubsection{Particles} \label{sec:MPI-particles} In a classical mechanical model, `particle' refers \emph{metaphysically} to an indefinitely persistent, continuously point-localized object which bears state-dependent and state-independent~(`intrinsic') properties, and whose position can be precisely measured by an ideal observer without affecting its properties.   The \emph{operational} meaning of `particle' hinges upon standardized procedures that empirically ground sufficiently many of these metaphysical assertions to a sufficient degree.  For example, there are established procedures for approximately measuring the position of a particle~(or centre of mass of particle-like physical object).  There are also procedures for measuring the other properties ascribed to a particle~(particularly its intrinsic properties) on the basis of successive position measurements, which procedures depend, in turn, upon procedures for \emph{reidentifying} a given particle in a system of interacting particles~(reidentifiability being the counterpart of persistence)\footnote{Some of these procedures will be discussed in Sec.~\ref{sec:classical-reidentification}.}.

Hence, in a classical particle model, an index~$i$ labels a persistent and reidentifiable particle, which reflects both metaphysical and operational meanings of `particle'.   However, the metaphysical meaning invariably outstrips what can, \emph{in practice}, be operationally grounded and empirically tested.  For example, observations of an object's position are necessarily imprecise and are performed at discrete moments of time, which evidently cannot \emph{fully} ground the metaphysical claim that the object is continuously point-localized.  Nevertheless, such metaphysical assertions are generally retained unless there are compelling reasons to do otherwise. 

In the foil quantum model of Step I, in the quantum state~$\psi(x_1, x_2, \dots, x_n)$, the symbol~$x_i$ is ordinarily assumed to retain its classical meaning of referring to the position of particle~$i$.  It is widely accepted that, due to the active nature of quantum measurement, it is unsafe to assume that~$x_i$ refers to directly to the current location of a particle, but rather refers to the value of position that is obtained upon observation.  Correspondingly, at the metaphysical level, it underdetermined whether or not `particle' refers to a physical object that manifests at a point upon position measurement, rather than one that is continuously point-localized even in the absence of measurement.  

However, the reading of index~$i$ as a \emph{particle label} also reflects an assumption of particle persistence, a reading that is inherited from standard practice in the quantum modelling of systems of \emph{non-identical} particles.  In the QSA, \emph{this reading is physically necessary} in Step II, where NOR requires that one determine whether the wavefunctions~(in general, the reduced states) of two or more particles overlap.

It is commonplace to \emph{extend} the particle-label reading of index~$i$ to the model constructed in Step III.  That is, the index~$i$ in \emph{symmetrized states} such as~$\psi(x_1, x_2) = [\alpha(x_1)\beta(x_2) \pm \beta(x_1)\alpha(x_2)]/\sqrt{2}$ is routinely assumed to refer to particle~$i$.  To ascertain whether such a reading is appropriate, note that, in Step III, the SOC restricts observables to those that are symmetric in the indices, so that the formalism itself prohibits description of a measurement performed upon particle~$i$.   Thus, in Step~III, the customary physical meaning ascribed to index~$i$ is untested by applications of the QSA, and thus is \emph{not} part of the MPI.  However, it is far from clear from inspection of the QSA to what~(if anything) the indices \emph{do} refer. 

As mentioned in the Introduction, some authors have asserted that the PLI should be dropped~\cite{DieksLubberdink2011,Caulton2014}.  However, these authors do not take into account the entire QSA, and accordingly do not observe that the PLI is needed in both Step I (to construct a foil model) and Step II (in order to apply NOR), so that dropping the PLI \emph{entirely} is not a viable option.  However, retention of the PLI in Steps~I and~II only to drop it in Step~III is unappealing unless one has a clear \emph{a priori} justification for \emph{changing} ones reading mid-stream.

\subsubsection{Systems}\label{sec:system-meaning}

In classical particle mechanics, a `system' commonly refers to an idealized collection of mass points that is not subject to external forces.   In a quantum model, the formalism's sharp distinction between `system' and `measurement' introduces ambiguity into the referent of `system', which is one facet of the so-called quantum measurement problem.  

In a quantum model of a system of identical particles, the notion of `system' is problematic in an additional way.  As mentioned above, Dirac's original presentation of the rules for handling identical particles  presumes that the user of these rules has already determined what `system' refers to.  This is, however, a consequential choice.   For example, consider two positrons, each confined to a separate trap~(modelled as a near-infinite potential well).  If one treats each positron as a system in its own right, then the quantum model of each would be obtained by solving the appropriate single-particle wave equation, and the state space of the composite model would be the tensor product of the single-particle Hilbert spaces.  However, if the `system' boundary is stretched to encompass \emph{both} positrons, then the SOC would prohibit the operator~$\hat{x}_1$, so that there would be no operator that describes a position measurement performed upon one of the positrons, despite their ostensible separation\footnote{In Step III of the QSA, if one drops the SOC---\emph{i.e.} one admits single-particle measurement operators---whilst retaining the SP,  then the expected position,~$\ev{\vect{r}_i}$ of positron~$i$ is \emph{mid-way} between the traps, which is contrary to physical intuition.}.

The \emph{no-overlap rule}~(NOR) resolves the above situation:~in the foil model, there is essentially no overlap of the states of the two positrons, hence the symmetrization constraints should not be applied.  In practice, use of this rule is ubiquitous.  For example, physicists commonly say that two photons are `distinguishable' if they are in different optical modes~(and accordingly their state can be rendered in the state formalism without carrying out symmetrization), and this interpretation underpins everyday use of the Fock space formalism.
Hence, it is essential that NOR be taken into account in any metaphysical interpretation of the QSA.  However, it is far from clear how to reconcile it with any standard interpretation of the SP.

\subsection{Interpretation}
\label{sec:MI}

As discussed earlier, claims about the nature of identical particles which are not part of the QSA and the MPI belong to the realm of metaphysical interpretation.  Ideally, a metaphysical interpretation \begin{inlinelist} \item provides a `metaphysical profile' of identical quantum particles, which clearly contrasts such particles with those familiar from classical physics; and \item traces this profile securely back to the QSA or, even better, to elementary facts upon which quantum theory or the QSA are based. \end{inlinelist} 

The interpretive challenge, however, is immense:~as we have discussed, the QSA is an amalgam of distinct rules with their own rationales, which makes it difficult to consistently interpret the MPI.

The two most prominent interpretational streams are based around the concepts of indistinguishability and (non-)individuality.  As indicated earlier, the first originates in Dirac's claim that identical particles in a system are `absolutely indistinguishable'\footnote{Here, `absolutely' means `for all observers, even ideal observers'.  This contrasts with Gibbs' indistinguishability, which is true only for limited observers.}, while the second originates in Bose's algorithm for quantum statistics~\cite{Bose1924}, which has been interpreted to imply a loss~(total or partial) of `individuality', this also being a common interpretation of the Fock formalism.  Both streams have lead to a voluminous literature; we restrict ourselves to a few brief remarks. 

As discussed above, the notion of indistinguishability is not part of the QSA and MPI, and its relation to these is tenuous.  Most importantly, indistinguishability is often taken as \emph{justification} of all or part of the QSA.   If this were true---that is, if indistinguishability were a \emph{general physical principle} that leads inexorably to the QSA~(or the most important parts thereof)---then it would indeed be appropriate to use it as a basis for interpretation of the QSA.  However, this is far from the case:~indistinguishability does not account for the SP%
\footnote{See Dirac's comments in~\cite[\S62~(p. 201)]{Dirac30} and~\cite[\S54~(p. 211)]{Dirac47}, and subsequent analyses of common arguments that purport to derive the SP~\cite{MG64, Girardeau65}.}, does not account for SOC unless one makes assumptions that go beyond the MPI~(see Remark~\ref{rem:justification-of-SOC}), and does not account for NOR. 

The non-individuality stream rests upon a reading of the Fock space formalism, and traces back to the fact that, unlike the QSA, this formalism eschews indices that can be readily interpreted as particle labels.  However, for fixed particle numbers, the Fock space notation is simply a compact rewriting of symmetrized states~\cite{Fock1932,Gordon2003}, and thus reflects only part of the QSA.  In particular, the non-individuality reading neglects both NOR and the construction of the foil model.  As mentioned in the Introduction, a few authors have recognized that some kind of `restricted individuality' is required to make sense of such phenomena as particle tracks~(which corresponds to NOR)~(see e.g.~\cite[p.~114-115]{Schroedinger1950} and~\cite[\S5]{Bigaj2020b}), but no explicit metaphysical profile of objects of such restricted individuality has thus far been proposed.

\section{\label{sec:persistence-and-reidentification}Persistence and Reidentification in Classical and Quantum Physics}

In this section, we begin our construction of a new interpretation of identical quantum particles.  For the reasons described in Sec.~\ref{sec:QSA-and-difficulties}, we do \emph{not} base this interpretation on the QSA.  Rather, we begin by reflecting on the physical meaning of `particle' or `physical object' in the context of classical mechanics and quantum mechanics.  Our focus is on the relationship between the notion of persistence---a metaphysical notion---and its operational counterpart, namely reidentification.  Our goal is to investigate to what extent the assumption of object persistence is undergirded by operational procedures for object reidentification in four cases:~\begin{inlinelist} \item solid objects in everyday life; \item classical particle mechanics; \item quantum systems of nonidentical particles; and \item quantum systems of identical particles. \end{inlinelist}  %
 
As we shall see, as a consequence of the abstraction and quantification of physical bodies carried out in the process of constructing physical theories~(for the purpose of formulating precise predictive models), the resources available for reidentifying bodies has been gradually eroded in moving from (i) to~(iv)---see the summary in Tables~\ref{tbl:reidentification1} and~\ref{tbl:reidentification2}.  Our conclusion is that, in case~(iv), the erosion is such that the assumption of \emph{unconditional} persistence is deprived of empirical cover.

\subsection{Everyday life.}
In everyday life, we are scarcely aware of the distinction between persistence and reidentifiability.  The visual and tactile appearances of naturally-occurring physical bodies are so rich and diverse that they yield an essentially unique perceptual signature which usually suffices for their reidentification.  In those rare circumstances when one is confronted with many objects of similar appearance, one can fashion perceptual handles by various means---using a microscope to see the distinctive pattern of nicks in several similar coins, or affixing a tiny barcode tag onto each of the ants in a colony.  Alternatively, one can reidentify an object---such as a particular bird in a flock as it passes overhead---by tracking it intently over time\footnote{When considering a composite object over longer time-scales, during which its internal composition can change, its reidentification depends upon a criterion of identity.  In the well-worn example of the ship of Theseus, one must decide whether a ship is considered `the same' once some or all of its planks have been replaced by ones that have the same \emph{form} as the original ones.  Reidentification of an organism through its lifespan---\eg a bird from single-cell to adult---constitutes an even more challenging example of this type, where the identity criteria is typically weakened to require only \emph{continuous change} in form.}.

\linespread{1.30}
\addtolength{\tabcolsep}{5pt}
\begingroup
\begin{table*}[!h]
\centering
	\begin{ruledtabular}
    \begin{tabular}{p{3.90cm}p{1.60cm}p{4.30cm}p{4.00cm}} 
    {\bf } 	 \tableheadtext{Physical situation} 	
    																	& \tableheadtext{Momentary \newline Appearance}   
    																	& \tableheadtext{Primary Means of \newline Reidentification}  
																	& \tableheadtext{Secondary Means of \newline Reidentification} 
																	\\ \hline

\smallskip\tabletext{Everyday objects}   	
																	& \smallskip\tabletext{Rich, detailed}   	
																	& \smallskip\tabletext{Similarity of momentary appearances \emph{(e.g. people in a small group)}}
																	& \smallskip\tabletext{Tracking through time \newline \emph{(e.g. flock of birds overhead)}} 
																	\\      
\smallskip\tabletext{Single classical particle}   	
																	& \smallskip\tabletext{Point-like events}   	
																	& \smallskip\tabletext{Track continuously \emph{(enables measurement of state-independent properties)}}
																	& \smallskip\tabletext{None} 
																	\\
																	
\smallskip\tabletext{Several nonidentical classical particles}   	
																	& \smallskip\tabletext{Point-like events}   	
																	& \smallskip\tabletext{Track each continuously through arena of interaction} 
																	& \smallskip\tabletext{Measure state-independent properties of each before and after interaction phase}
																	\\
						
\smallskip\tabletext{Several identical classical particles}   	
																	& \smallskip\tabletext{Point-like events}   	
																	& \smallskip\tabletext{Track each continuously through arena of interaction}
																	& \smallskip\tabletext{None} 
																	\\
	\\
    \end{tabular}
	\end{ruledtabular}
\caption{\label{tbl:reidentification1} \emph{Reidentification in everyday experience and in classical mechanics.}}
\end{table*}
\endgroup
\linespread{1.618}

\subsection{Classical Mechanics}
\label{sec:classical-reidentification}
Classical physics exalts our most readily quantifiable sense---the visual---over the others, resulting in the Cartesian conception of physical bodies as geometric entities of pure extension.  Once extension is dropped as a primitive property, one arrives at the Newtonian conception of a physical body as a collection of interacting \emph{particles}---extension-free persistent entities embedded in a geometric space which serve as centers of force.  

\begin{figure}[!h]
\centering
\vspace{2em}
     \begin{subfigure}[b]{0.48\textwidth}
         \centering
         \includegraphics[width=\textwidth]{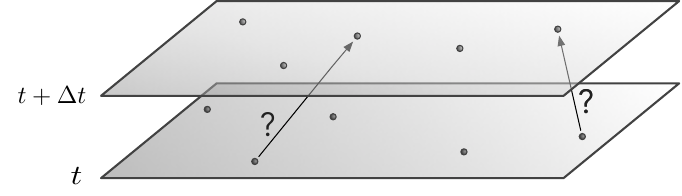}
         \caption{\label{fig:reidentifying-classical-particle-pane1}}  
     \end{subfigure}
     \hfill
     \begin{subfigure}[b]{0.48\textwidth}
         \centering
         \includegraphics[width=\textwidth]{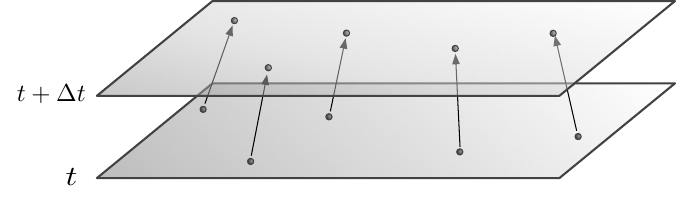}
         \caption{\label{fig:reidentify-classical-particles-pane2}}
     \end{subfigure}
\linespread{1.3}
\caption{\label{fig:reidentifying-classical-particles}\emph{Reidentification of classical particles.}  An ideal experimenter records a pattern of flashes at closely-separated times,~$t$ and~$t + \Delta t$, which are assumed to be generated by persistent particles.  (a)~If the experimenter cannot assume that the particles move continuously~(or via small `jumps'), he cannot reidentify them---he is unable to say `\emph{this} flash at~$t$ is generated by the same particle as \emph{that} flash at~$t + \Delta t$', despite knowing~(from the assumption of persistent particles) that the same particle is responsible for one of the flashes at~$t$ and one of the flashes at~$t + \Delta t$.  (b)~If the particles move continuously~(or via small `jumps'), then approximate reidentification becomes possible.  If the particles move continuously~(as posited by classical physics), the precision of such reidentification can theoretically be increased indefinitely by reducing the size of~$\Delta t$, and tends to exactness in the limit at~$\Delta t \rightarrow 0$.}
\end{figure}

Position is the only directly measurable property of a Newtonian particle, assumed to be rendered perceptible to any degree of precision via our visual sense~(if appropriately instrumentally augmented).  Therefore, unlike the objects of everyday experience, all Newtonian particles instantaneously appear alike---as a point-like spatial event or \emph{flash}~(see Table~\ref{tbl:reidentification1}).   Accordingly, \emph{tracking} is the primary means of reidentifying such particles, the possibility of which is granted by theoretical assumptions:~each particle is assumed to move continuously in space, and an ideal experimenter is assumed to be capable of making precise measurements of a particle's position without disturbing any of its properties~(see Fig.~\ref{fig:reidentifying-classical-particles}).  We refer to this means of reidentification as $t$-reidentification\footnote{Strictly speaking, $t$-reidentification requires that the space be endowed with a topology, not a metric.  Operationally, the ideal experimenter need only be able to \emph{precisely locate} the particle at each moment in time~(figuratively, to \emph{drop a marker} at the particle's location), but need not have any means to \emph{quantify} its location~(\ie assign coordinates) relative to a reference system.  Such an experimenter must be equipped with a clock that faithfully tracks the forward flow of time, but the metric over time is irrelevant.}.   

Since particles instantaneously appear alike, the state-independent properties---mass and charge---attributed to a particle by the theories of classical physics are concealed behind a featureless exterior, only becoming manifest~(and thus measurable) when the particle moves in an experimental context under an experimenter's control.  Thus, the measurement of the state-independent properties of a particle, which are theoretically attributed to the particle in the moment, require that it be reidentifiable over an interval of time. 

\begin{figure}[!h]
     \begin{subfigure}[b]{0.45\textwidth}
         \centering
         \includegraphics[width=\textwidth]{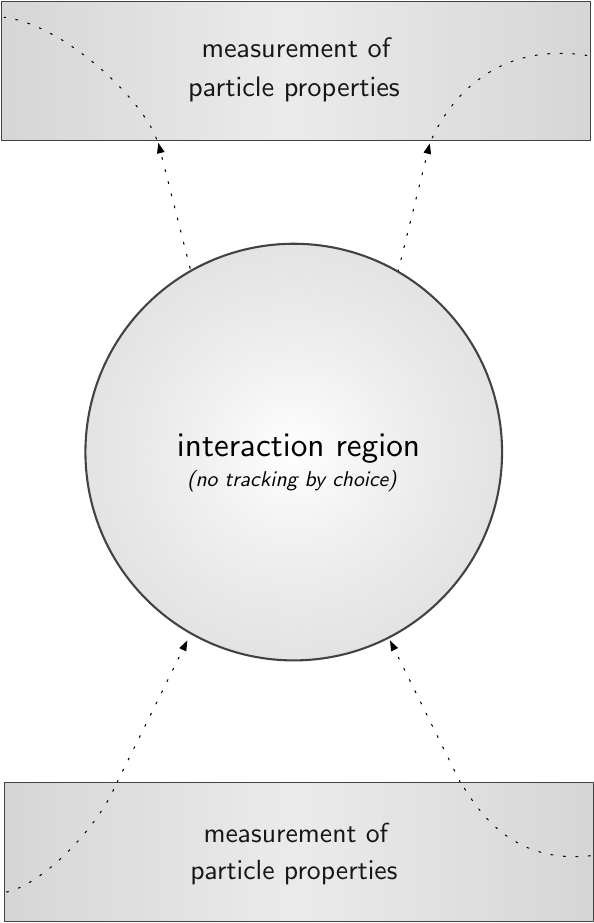}
         \caption{\label{fig:reidentifying-interacting-classical-particles-pane1}}  
     \end{subfigure}
     \hfill
     \begin{subfigure}[b]{0.45\textwidth}
         \centering
         \includegraphics[width=\textwidth]{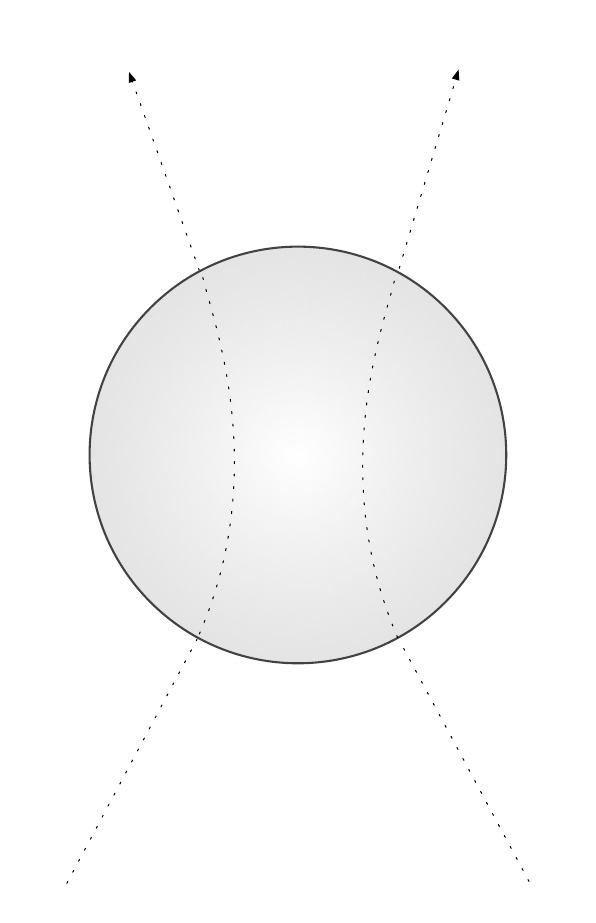}
         \caption{\label{fig:reidentifying-interacting-classical-particles-pane2}}
     \end{subfigure}
\linespread{1.3}
\caption{\label{fig:reidentifying-interacting-classical-particles}\emph{Reidentifying interacting classical particles.} (a)~Nonidentical particles can be reidentified by measuring their state-independent properties~(mass, charge) before and after the interaction; tracking within the interaction region is unnecessary.  (b)~Reidentification of identical particles requires that they be tracked sufficiently precisely as they pass through the arena of interaction.}
\end{figure}

Once the state-independent properties of a set of particles are so measured, they can sometimes serve as a \emph{secondary} means of reidentification.  In particular, given a set of non-identical particles, a specific particle can be reidentified after its interaction with the others provided that the particles' state-independent properties are measured before and after such an interaction~(see Fig.~\ref{fig:reidentifying-interacting-classical-particles-pane1}).  We refer to such a means of reidentification as $p$-reidentification.

However, given a set of \emph{identical} particles~(a theoretical possibility that is granted by classical mechanics), $p$-reidentification is evidently impossible\footnote{We emphasize that $p$-reidentification refers to reidentification through measurement only of \emph{state-independent} properties, and its significance is that it is the only property-based means of reidentification that is independent of the experimental context.  Reidentification of identical classical particles is possible in those \emph{special cases} in which one can reidentify through measurements of the particles' \emph{state-dependent} properties.  For example, two charged particles interacting electromagnetically but trapped on different sides of a partition can be reidentified by their location.}.  Accordingly, reidentification of a specific particle can, in general, \emph{only} be achieved by $t$-reidentification~(see Fig.~\ref{fig:reidentifying-interacting-classical-particles-pane2}).  

\subsection{Quantum mechanics}
\label{sec:quantum-reidentification}

Quantum theory dispenses with many of the key conceptions that are central to classical physics.  In particular, the quantum formalism dispenses with the notion that measurement is a passive process that can register the position of a physical object without affecting its physical state.  Instead, measurement is regarded an \emph{active process} whose outcome can only be predicted probabilistically, and which leads to a corresponding change in the object's state.  Consequently, in the \emph{design of any experiment,} the free evolution of a physical object, or its interaction with others, both of which are treated by the formalism as a deterministic process, must be segregated from any measurement processes if they are to be studied \emph{as is}. 

As we shall detail below, this design constraint erodes the resources available for particle reidentification, even by an ideal observer~(see Table~\ref{tbl:reidentification2}). 

\linespread{1.30}
\begingroup
\begin{table*}[!h]
\centering
	\begin{ruledtabular}
    \begin{tabular}{p{3.00cm}p{1.60cm}p{4.50cm}p{4.50cm}} 
    {\bf } 	 \tableheadtext{Physical situation} 	
    																	& \tableheadtext{Momentary \newline Appearance}   
    																	& \tableheadtext{Primary Means of \newline Reidentification}  
																	& \tableheadtext{Secondary Means of \newline Reidentification} 
																	\\ \hline

\smallskip\tabletext{Single quantum \newline particle}   	
																	& \smallskip\tabletext{Point-like events}   	
																	& \smallskip\tabletext{Inferences based on pre- and post-interaction measurements~\emph{(unable to track during interaction phase)}}																												& \smallskip\tabletext{None} 
																	\\														
										
\smallskip\tabletext{Several nonidentical \newline quantum particles}   	
																	& \smallskip\tabletext{Point-like events}   	
																	& \smallskip\tabletext{Measure particles' state-independent properties before and after interaction phase}
																	& \smallskip\tabletext{None~\emph{(unable to track during interaction phase)}} 
																	\\			
																									
\smallskip\tabletext{Several identical \newline quantum particles}   	
																	& \smallskip\tabletext{Point-like events}   	
																	& \smallskip\tabletext{Approximate tracking possible in some contexts \emph{(bubble chamber, with separated tracks)}, not
others \emph{(electrons in a helium atom)}}
																	& \smallskip\tabletext{None~\emph{(unable to track during interaction phase)}} 
																	\\		
	\\
    \end{tabular}
	\end{ruledtabular}
\caption{\label{tbl:reidentification2} \emph{Reidentification in quantum physics.}}
\end{table*}
\endgroup
\linespread{1.618}

\subsubsection{Reidentification in a one-particle quantum system}
\label{sec:reidentification-one-particle}
In its minimal instrumental interpretation, the quantum formalism takes as primitive the notions of \emph{physical system}, \emph{measurement}, and \emph{interaction}, and applies to an idealized experiment in which the system is subjected to a sequence of measurements and interactions implemented by macroscopic devices\footnote{For our present purposes, the question of what kinds of physical processes constitutes a measurement is irrelevant---it suffices that there are experimental procedures that are universally accepted as instantiating measurement processes.}.  Thus, the physical system is treated as a persistent entity, retaining its identity for the duration of the experiment.  Measurements differ from interactions in that measurements yield macroscopically-observable events~(such as scintillations on a phosphorescent screen or audible clicks).   It is through these events that the physical system comes to be indirectly known.  

For example, consider an experiment that an experimenter would typically describe as one in which electrons are liberated at a heated filament, accelerate through a wire-loop detector, diffract through a crystal lattice, and then impact a phosphorescent screen.  The experimenter presumes that a click of the wire-loop detector followed moments later by a point-like flash on the screen both constitute imprecise position measurements of the same electron.  The interaction of the electron with the crystal lattice is then modelled via a potential function typically drawn from a classical physical description of the crystal.  

In such an arrangement, the experimenter's confidence that he has \emph{reidentified} an electron---that the entity which presently generates a scintillation is the same as the entity that elicited a click moments earlier---does not rest upon having tracked that electron from filament to screen.  Indeed, due to the invasive nature of quantum measurements, any attempt to carry out fine-grained tracking of the electron would uncontrollably interfere with the process during which the electron interacts with the crystal.  Thus, reidentification necessarily depends upon various \emph{indirect}, empirically-grounded inferences:~the scintillations practically cease once the filament is cooled to room temperature; the rate of scintillations increases with filament temperature; the scintillation pattern is shifted if an external magnetic field is applied~(as would be expected for charged particles emitted from the filament); and so forth.

\begin{figure}[!h]\linespread{1.3}
\begin{center}
\includegraphics[width=1.0\textwidth]{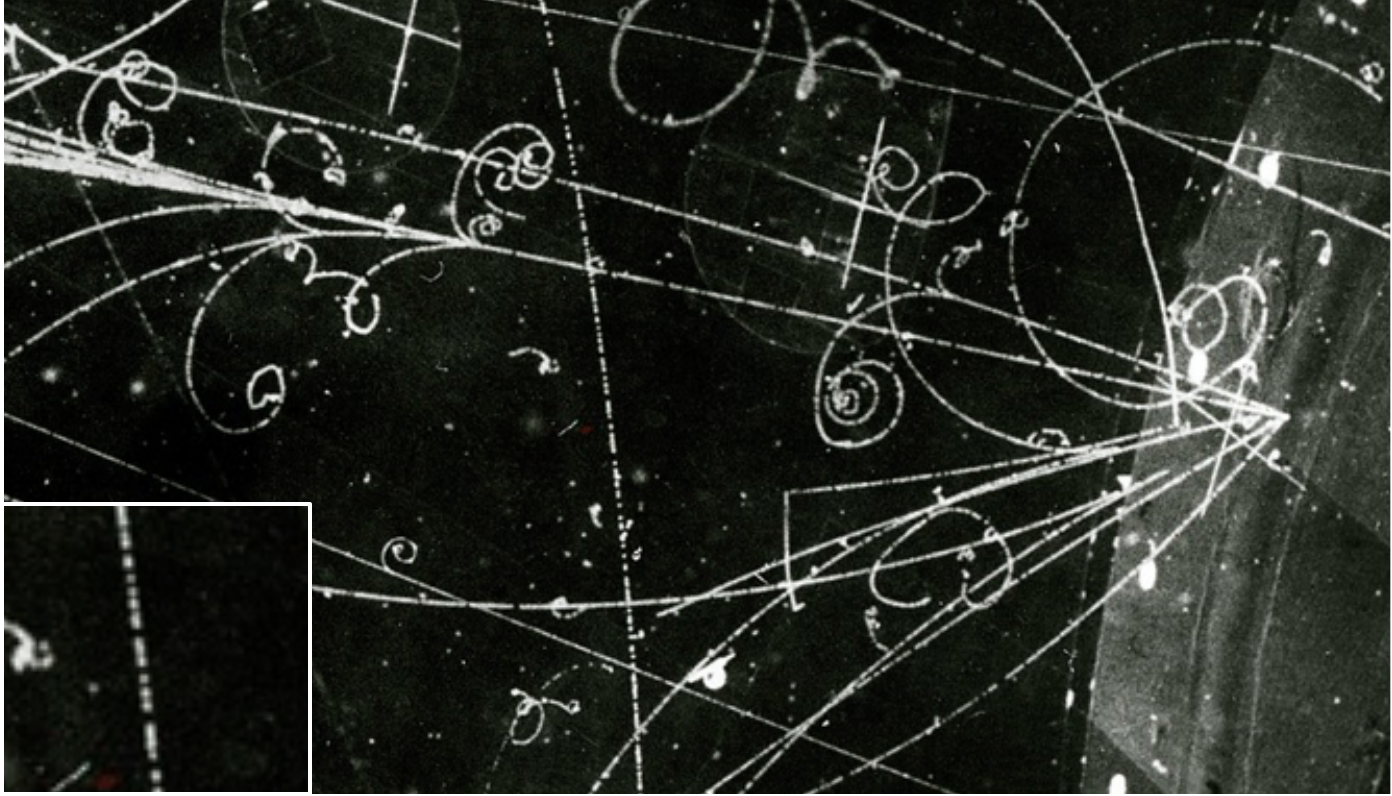}
\caption{\label{fig:bubble-chamber}\emph{Reidentification and measurement of a particle's state-independent properties in a bubble chamber.}  A bubble chamber image is parsed into distinct `tracks', each of which is assumed to be due to a specific particle.  The inset~(bottom left) indicates that a track in fact consists of a sequence of discrete bubbles.   The particles move in helical trajectories due to an applied magnetic field, which enables calculation of the particles' charge to mass ratio. \emph{(Image courtesy of Brookhaven National Laboratory)}.
}
\end{center}
\end{figure}

The measurement of a particle's state-independent properties depends upon the possibility of fashioning an experimental context in which the experimenter can reliably infer that successive measurement outcomes are generated by the same particle, despite being unable to continuously track the particle.  Such a situation obtains in a bubble chamber:~an experimentalist parses a bubble-chamber image into \emph{tracks}, each presumed to have been generated by a single particle during a portion of its life-history~(see Fig.~\ref{fig:bubble-chamber}).   Each track consists of a sequence of separated yet closely-spaced bubbles, and these bubbles are modelled as the outcomes of successive inexact position measurement of the same object, between which the particle is assumed to evolve through interaction with its environment.  In certain portions of a track, the particle can be safely assumed to be under the influence of an environment~(such as an arrangement of electric and magnetic fields) that is under the experimenter's control, enabling its state-independent properties to be measured.

\subsubsection{Reidentification in a system of two particles}

Consider an experiment to study the interactions of two \emph{non-identical} particles.  Let us suppose that, in the first phase of such an experiment, one has measured the state-independent properties of each particle by passing each through a bubble chamber.   Now let the particles be allowed to interact with one another, without the intrusion of measurement.  Once the interaction phase is completed, the particles' state-independent properties can again be measured.  As the particles are non-identical, they can, in principle, be perfectly reidentified~(via $p$-reidentification).

However, consider an analogous experiment to study the interaction of two \emph{identical} particles, such as two electrons.  Reidentification of each particle by $p$-reidentification is no longer possible.  But, unlike the corresponding classical experiment, \emph{particle tracking during the interaction phase is also impermissible} if one wishes to study the undisturbed interaction of the particles.  Therefore, the experimenter no longer possesses any reliable means to reidentify the particles for \emph{arbitrary} interactions.  As in the above-considered case of two identical classical particles, the experimenter can exactly or probabilistically reidentify in special circumstances.  For example, in the extreme case where the electrons happen to be confined to separate electromagnetic traps in different labs, one can safely assume that the pre- and post-detections in a given lab are generated by the same electron.  Similarly, if two electrons are prepared in orthogonal spin states, and subsequently interact only through their spatial degrees of freedom~(\ie without involving their spin degrees of freedom), they can be reidentified through post-interaction measurement of spin states~(see~\cite[\S3.4]{Feynman-v3} and~\cite[\S4]{Bigaj2020}).    However, for arbitrary interactions, such as a sufficiently energetic collision between two electrons, no state-dependent properties are available for reidentification.  

\section{Persistence and Nonpersistence Models of Flashes}
\label{sec:persistence-nonpersistence}

As we have seen above~(Sec.~\ref{sec:persistence-and-reidentification}), in an experiment to study the interaction of two identical quantum particles, reidentification is not, in general, possible.  \emph{Is one still justified in assuming that, during the interaction phase, there exist two individually persistent entities?}  

Since we have lost any reliable means of reidentification while two electrons are interacting, and since this loss of capability is due to the confluence of two fundamental assumptions---the existence of identical particles in nature, and the active nature of quantum measurements---the assumption of \emph{individual persistence} during this interaction phase is metaphysically exposed.  
Ontological modesty beckons us to contemplate some way of conceiving what happens during the interaction phase which does not commit us to this assumption.  

If we wish to entertain a kind of context-dependent persistence, we seem to have no choice but to give up the idea of `particles' as basic.  Seemingly the only recourse is to \emph{take the flashes themselves as basic}, and thus to regard the notion of particle as a \emph{secondary} notion---as a conceptual device for threading a sequence of flashes, a device that is only applicable in certain experimental contexts.

\begin{figure}[!h] \linespread{1.3}
\begin{center}
     \begin{subfigure}[b]{0.45\textwidth}
         \centering
         \includegraphics[width=\textwidth]{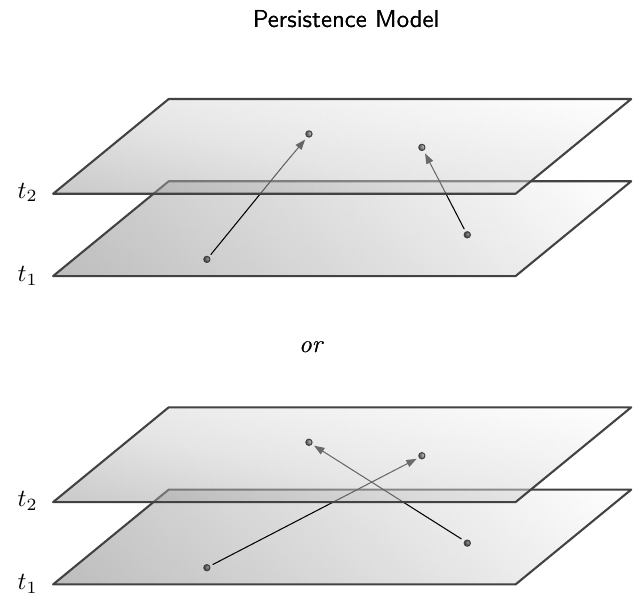}
         \caption{\label{fig:persistence-and-nonpersistence-models-pane1}}  
     \end{subfigure}
     \hfill
     \begin{subfigure}[b]{0.45\textwidth}
         \centering
         \raisebox{0.5\height}{\includegraphics[width=\textwidth]{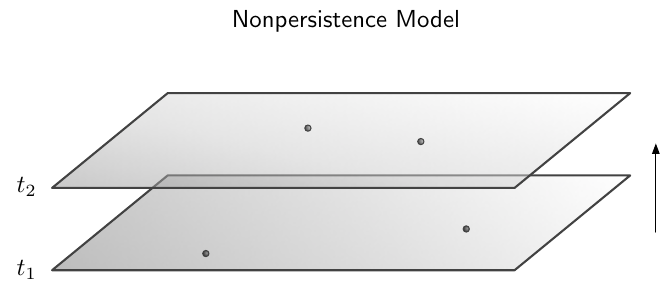}} 
         \caption{\label{fig:persistence-and-nonpersistence-models-pane2}}
     \end{subfigure}
\caption{\label{fig:persistence-and-nonpersistence-models}\emph{Persistence and nonpersistence models of flash-data.} Two flashes are registered at time~$t_1$ and at time~$t_2$. Two different object-models of this data are possible.  (a)~The persistence model posits that each flash is the momentary appearance of a persistent object~(`particle').  Hence, according to the persistence model, two possible transitions are compatible with the flash data, only one of which occurred.   (b)~The nonpersistence model posits that both flashes at each time are the momentary appearance of a single holistic object.} 
\end{center}
\end{figure}

This shift of ontological ground---away from \emph{object}, towards \emph{event}---affords the flexibility to construct two \emph{different} object-models of the same flash-data~(see Fig.~\ref{fig:persistence-and-nonpersistence-models}).
Consider an experiment in which two flashes are registered at~$t_1$ and then later at~$t_2$, and let us refrain from specifying the experimental context~(one may have two `particles' in a bubble chamber, two interacting `particles', or some other context).  The first model---the \emph{persistence} model---is the familiar one:~each flash is presumed to be the momentary appearance of a persistent object~(`particle'), which enables one to say that one of two possible \emph{transitions}\footnote{Note that quantum theory does \emph{not} give us warrant to say that the particles traversed specific paths between flashes.} occurred unseen between times~$t_1$ and~$t_2$~(see Fig.~\ref{fig:persistence-and-nonpersistence-models-pane1}).  That is, the persistence model gives us warrant to say that one of the flashes at~$t_2$ was generated by \emph{the same} entity as one of the flashes at~$t_1$.

However, a second model---the \emph{nonpersistence} model---of the same flash-data is also possible.  In that model, one regards the two flashes at each time as the momentary appearance of a \emph{single} persistent object\footnote{For the avoidance of possible terminological confusion: despite its name, the nonpersistence model \emph{does} presume that there exists a persistent entity in the time interval of its application.  More broadly speaking, any application of the Feynman formalism in the interval~$[t_1, t_2]$ presumes that there exists a persistent system during this interval.  Both the persistence and nonpersistence models are compatible with this presumption.  They differ, however, in what they assert about the nature of that system. The persistence model posits that the system consists of \emph{two}~(or, in general,~$N$) persistent entities; the nonpersistence model that the system consists of a \emph{single} persistent entity.}~(see Fig.~\ref{fig:persistence-and-nonpersistence-models-pane2}).  What speaks against such a posit is our experience with everyday physical objects, which habituates us to believe that the location of physical objects coincides with the locations of their appearances.  But, if we are prepared to set aside this aspect of mental conditioning and entertain a more abstract conception of physical object, we see that such a model indeed avoids positing the existence of individually persistent objects in contexts where this posit lacks empirical warrant.

\subsection{\label{sec:reconstruction}Derivation of the Feynman's quantum symmetrization algorithm for a system of identical particles}

Above, we have proposed two quite different object-models of the same event-data.  But how can one use \emph{two} models of the same data to construct a predictive theory?  In particular, while the persistence model effectively parses the data into `tracks' underpinned by two persistent objects, the nonpersistence model offers no such analysis, and might well for this reason alone be dismissed as not providing the traction necessary to build up a predictive theory.

However, rather than attempting to build a theory on the basis of either model alone, perhaps it is possible to somehow \emph{synthesize} them.  But how it could it be possible to do so when these models are \emph{inconsistent} in their posits as to the nature of the entity~(or entities) that underpin the data?  The key point is that although the two models differ in the claims that they make about the nature of the entity~(or entities) that exist in between the detections, these claims can, by their very nature, only be \emph{indirectly} probed via experiment, for the experimenter actually \emph{observes} detector clicks or flashes, not the posited entities in themselves. 
Thus, the data provides sufficient latitude to combine key features of \emph{both} models into a predictive model. 

\begin{figure}[!h] \linespread{1.3}
\begin{center}
\includegraphics[width=0.98\textwidth]{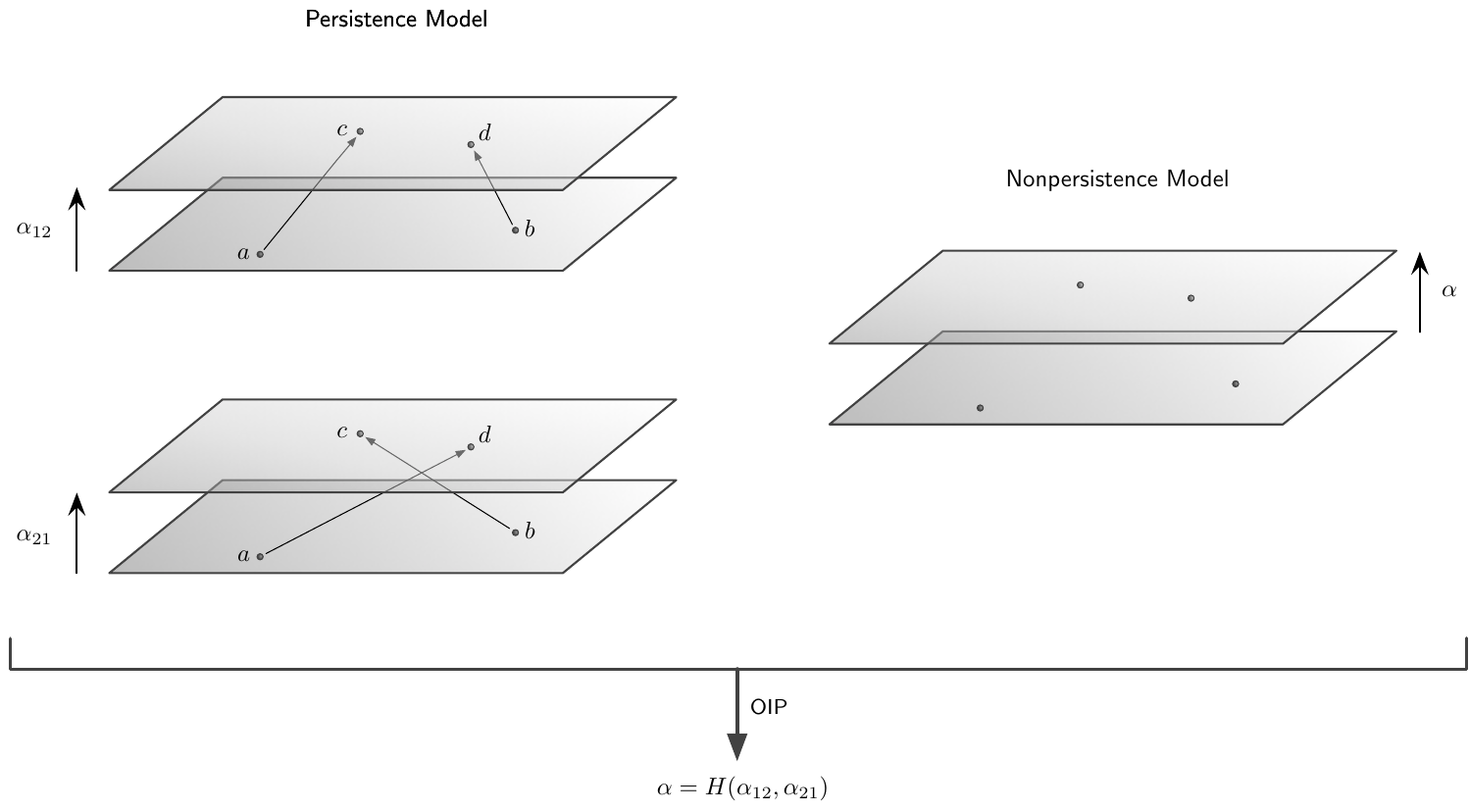}
\caption{\label{fig:model-synthesis}\emph{Synthesis of persistence and nonpersistence models.} An ideal experimenter registers two flashes at~$t_1$ and two flashes at~$t_2$. Two models---a persistence model and a nonpersistence model---of this data are constructed, described in the Feynman formalism, and then synthesized by the operational indistinguishability postulate~(OIP). } 
\end{center}
\end{figure}

In~\cite{Goyal2015, Goyal2019a}, it is shown how these models can be mathematically synthesized.  Here we summarize the main features of the derivation and their significance.
The synthesis takes place in the context of the Feynman formulation of quantum theory\footnote{The Feynman formalism is described in~\cite{Feynman49}, and is systematically derived and restated in an operational framework in~\cite{GKS-PRA}.}, within which the persistence and nonpersistence models can readily be described~(see Fig.~\ref{fig:model-synthesis}).
Specifically, in the persistence model, one treats the two flashes registered at~$t_1$ as two distinct outcomes, each the manifestation of a measurement on a separate persistent subsystem.  Accordingly, given the outcomes~$a$ and~$b$ registered at~$t_1$ and the outcomes~$c$ and~$d$ registered at~$t_2$, two possible transitions could have occurred---the transition~(which we shall arbitrarily call the \emph{direct} transition) in which the same subsystem is responsible for outcomes~$a$ and~$c$, and the \emph{indirect} transition in which the same subsystem is responsible for outcomes~$a$ and~$d$.  And, following the Feynman formalism, one assigns an amplitude to each transition, namely~$\alpha_{12}$ to the direct, $\alpha_{21}$ to the indirect.  In contrast, in the nonpersistence model, one treats the two flashes registered at each time as a \emph{single} outcome of a measurement performed on a physical system.  Accordingly, there is but a single transition between the outcome at~$t_1$ and the outcome at~$t_2$, which is assigned amplitude~$\alpha$.  

The synthesis of these models is achieved through the so-called \emph{operational indistinguishability postulate}~(OIP) which posits that~$\alpha = H(\alpha_{12}, \alpha_{21})$, where~$H$ is some function to be determined.
An analogous statement is posited for a three-stage experiment in which flashes are registered at \emph{three} successive times\footnote{In the case that flashes are obtained at times~$t_1, t_2$ and~$t_3$, then there are \emph{four} possible transitions, in the persistence model, compatible with this flash-data~(see Fig.~2(b) in~\cite{Goyal2015}). Denoting the corresponding transitions amplitudes~$\gamma_{11}, \gamma_{12}, \gamma_{21},$ and~$\gamma_{22}$, the OIP then stipulates that~$\gamma = \Gmat{\gamma_{11}}{\gamma_{12}}{\gamma_{21}}{\gamma_{22}}$, where~$\gamma$ is the transition amplitude in the nonpersistence model, and~$G$ is a function to be determined.}.

One can motivate the OIP by the idea that the actual transition amplitude should incorporate \emph{both} of the possible transitions posited in the persistence model, even though, according to the persistence model, only one such transition can occur.  That is, the persistence model should not be taken as `carving nature at the joints', but rather as an expedient way to analyse, and thus render tractable, the experimental data.  In a more general context where we suspect that this model is not strictly valid, the idea is that rather than dispensing with the model entirely, one can move beyond it in a manner that is strictly speaking at odds with the assumption of individual persistence that underwrites it by mathematically \emph{combining} the amplitudes of the two transitions that the model itself would deem mutually exclusive.

The form of the to-be-determined function~$H$ is importantly constrained in the following manner.  We have already noted that, when isolated from one another in a bubble chamber, two electrons \emph{can} be reidentified, and one therefore has reasonable grounds to regard the bubbles in each `track' as the manifestations of an individual persistent entity.   This idea motivates the \emph{isolation condition} which applies in the special case where one or more of the transition probabilities in the persistence model is zero.  In the case of two separated electrons in a bubble chamber, this would mean that one does not need to consider the possibility that the two electrons surreptitiously `swap' tracks.  In this case, the isolation condition posits that the persistence model alone should be adequate---that is, it is correct to say that each flash on a given track is due to a persistent individual.  This translates into a condition on~$H$, namely~$|H(z, 0)| = |z|$ and~$|H(0, z)| = |z|$.

From the assumptions described above\footnote{Specifically, the assumptions are: \begin{inlinelist} \item the OIP for two successive times,~$\alpha = H(\alpha_{12}, \alpha_{21})$, and three successive times~$\gamma = \Gmat{\gamma_{11}}{\gamma_{12}}{\gamma_{21}}{\gamma_{22}}$; and \item the isolation condition, $|H(z, 0)| = |z|$ and~$|H(0, z)| = |z|$. \end{inlinelist} }, the requirement of consistency\footnote{The requirement of consistency manifests as the requirement that, if there are two different ways to compute a transition amplitude, then these must agree.  Each such call for consistency yields a functional equation in terms of the unknown functions~$G$ and~$H$~(see Figs.~4--6 in~\cite{Goyal2015}).  One thereby obtains the following:~\begin{inlinelist} \item $\Gmat{\alpha_{12} \beta_{12}}{\alpha_{12}\beta_{21}}{\alpha_{21}\beta_{12}} {\alpha_{21}\beta_{21}}  = H (\alpha_{12}, \alpha_{21}) \, H(\beta_{12}, \beta_{21})$; \item $G\bigl( (\alpha_{12}+ \alpha_{12}')\beta_{12}, (\alpha_{12}+ \alpha_{12}')\beta_{21}, (\alpha_{21} + \alpha_{21}')\beta_{12}, (\alpha_{21} + \alpha_{21}')\beta_{21} \bigr) = H(\alpha_{12}, \alpha_{21}) \, H(\beta_{12}, \beta_{21}) + H(\alpha'_{12}, \alpha'_{21}) \, H(\beta_{12}, \beta_{21})$; \item $H^*(\alpha_{12}, \alpha_{21})  = H(\alpha_{12}^*, \alpha_{21}^*)$. \end{inlinelist}} %
 fixes the form of~$H$ for two `particles':~$H(\alpha_{12}, \alpha_{21}) = \alpha_{12} \pm \alpha_{21}$, where the~$\pm$ sign corresponds to bosonic~($+$) and fermionic~$(-)$ behavior, which is  Feynman's symmetrization postulate.  The derivation can also be extended to the more general case of~$N$ `particles'~\cite{Goyal2015}.  
As outlined in Appendix~\ref{sec:rQSP-and-interpretation}, these results can be re-expressed in terms of quantum states to enable comparison to the QSA.

\section{\label{sec:potential-parts}Identical quantum particles as potential parts}

The two doctrines of actual parts and potential parts differ in the ontological status they accord the parts into which a body can be divided.  The former posits that these parts exist independently from the body, so that the process of division simply \emph{reveals} them; the latter that these parts do not exist as individuals prior to division, so that division \emph{(co)-creates} them~\cite[Ch.~2]{Holden2006}.  Moreover, versions of the potential parts doctrine assert that the whole ceases to exist once the parts are actualized~\cite[Ch.~2, p.~12]{Digby1645}.  The actual parts doctrine is conceptually transparent---a body is simply an arrangement of actual parts, each of which can be characterised independently of their presence in the body---and is the basis of atomistic conception of classical physics.  In contrast, the potential parts doctrine, historically motivated in part by the rich part-whole relations characteristic of living organisms, has had little sway in natural science.  In this section, we argue that this doctrine is an illuminating lens through which to view a system of identical particles, and briefly discuss its implications for transtemporal identity.

\subsection{Interplay of actual and potential existence} 
As discussed in Sec.~\ref{sec:persistence-nonpersistence}, the notion of individual persistence is rendered context-dependent by taking detection events, rather than microscopic objects, as the primary reality.  This allows one to account for the same pattern of detection events via object-models---the persistence and nonpersistence models---that \emph{differ} in the microscopic objects that they posit as underpinning the event-data.  Accordingly, in certain situations---two electrons in separate traps, say---where the persistence model alone suffices, one can unequivocally say that each electron exists \emph{actually}, as actual parts of the two electron system.   

However, in other situations, such as two electrons in a helium atom, \emph{both} models are required to account for the system's behaviour.  In such situations, the models make contradictory assertions as to the extent objects:~one asserts the existence of two electrons, the other a single holistic object.  As both models play a non-negligible role in the empirical validity of the synthetic model, \emph{both} of their claims have warrant, yet neither has absolute validity.  Hence, it seems appropriate to say that the electrons exist \emph{potentially}, as potential parts of the system.  If the helium atom is completely ionized~(which is a process of division), the electrons \emph{emerge} as actual parts, while the holistic object posited by the nonpersistence model ceases to exist.

Thus we see that the interplay between the persistence and nonpersistence models appear to formally mirror the key aspects of the doctrine of potential parts.

\subsection{Restricted transtemporal Identity}
In the atomistic conception, the parts of a body or system exist continuously over a finite period of time.  The persistence of the parts underwrites their transtemporal identity:~it is a fact of the matter whether or not a given part at time~$t_2$ is the same as a given part at~$t_1$; and every part at~$t_2$ has a counterpart at~$t_1$.   In the potential parts doctrine, however, transtemporal identity can be lost since an object captured by---and integrated into---a whole suffers a loss of actuality.  

Consider the following example due to Lowe~\cite[p.62]{Lowe2001}~(also discussed in~\cite{Lowe1994}):
\begin{quote}
...if a single free electron, which we could label $a$, were to be captured by a helium ion---entering a state of entanglement with the single electron already present in the ion's shell---and subsequently a single electron, which we could label $b$, were to be released by the atom, then it seems that there would be no fact of the matter as to the truth or falsehood of the statement `$a$ is identical with $b$'.
\end{quote}
Our analysis bears out this assertion.  As discussed in Sec.~\ref{sec:persistence-nonpersistence}, the single free electrons---$a$ and $b$---are each well-described by the persistence model, and can be said to exist \emph{actually}.  Hence, before capture and after emission, the system consists of two actual parts~(an electron and a helium ion) which possess transtemporal identity~(and so can be meaningfully labelled).  However, upon capture by a helium ion, the identity of electron $a$ is submerged:~the helium atom requires both the persistence and nonpersistence models for its proper description, so electron $a$ exists potentially whilst part of the helium atom.  Due to the electron's submerged existence whilst in the atom, the thread connecting $a$ and $b$ is broken:~there is no fact of the matter as to whether $a$ is identical to $b$.

As this example illustrates, the loss of transtemporal identity is context-dependent.  Indeed, this loss is a matter of degree.  For example, if electrons $a$ and $b$ were---as viewed in the persistence model---to approach each other closely in a bubble chamber, so that there is a non-negligible probability of path cross-over~(\emph{i.e.} $\alpha_{21}$ cannot be neglected in comparison to~$\alpha_{12}$), then transtemporal identity would be \emph{partially}---rather than entirely---blunted.  Indeed, the values of~$\alpha_{12}, \alpha_{21}$ also provide a means to quantify the degree to which transtemporal identity is retained.

We thus see how it is possible to place Schroedinger's notion of `restricted individuality' on a clear conceptual foundation. 
\subsection{Boundary between systems of actual and potential parts}

The doctrine of potential parts raises some fundamental questions:~under what circumstances is it necessary to describe systems in nature in terms of potential parts, and when is it sufficient to conceive them in terms of actual parts?  For example, a pile of sticks can evidently be treated as a system of actual parts; and, more generally, the successes of Newtonian particle mechanics shows that this conception has wide applicability.  So, at what level of nature does thinking in terms of potential parts become necessary~(or illuminating), and what is gained in doing so?

Based on our analysis, we can offer an outline answer to these questions, at least insofar as physical~(as opposed to, say, biological) systems are concerned.   As we have seen in Sec.~\ref{sec:persistence-and-reidentification}, the peculiar behaviour of systems of identical quantum particles arises from the confluence of two key elements---\begin{inlinelist} \item  the particles are identical in their intrinsic properties; and \item measurement is an \emph{active} process. \end{inlinelist}  That is, the potential parts conception---at least in physics---becomes applicable under very special circumstances, being unnecessary either for systems of \emph{non-identical} quantum particles~(such as hydrogen atom), or for \emph{classical} systems of identical particles.   Moreover, even for a system of identical particles, if the isolation condition is met, the persistence model suffices, and the particles can be regarded as actual parts.

In addition, the transition to potential parts is associated with a new kind of holistic behaviour of the parts.  For example, in a helium atom, where~(as viewed through the lens of the persistence model) one would say `two electrons are interacting', the behavior of the helium atom~(as manifest, for example, in its atomic spectrum) cannot be deduced from this model alone, but additionally requires that one invoke the nonpersistence model.  The nonpersistence model could, accordingly, be viewed as the origin of a `higher-level law', and also as the origin~(via the process of model synthesis, formalized through the OIP) of the `downward causation' that (from the perspective of the persistence model) appears to act on the parts.   This suggests an intriguing connection with the biological whole-part relations that provide one historical motivation for the potential parts doctrine.

\section{\label{sec:conclusion}Concluding remarks}

We have presented a new conception of systems of identical quantum particles in which detection events are ontologically primary, and particles~(as persistent entities) are a secondary or emergent notion.   The need for such a radical departure from the atomistic conception originates in the confluence of two basic facts---\begin{inlinelist} \item the existence of particles that are \emph{entirely alike} in their intrinsic properties~(in contrast to the objects of everyday experience); and \item the \emph{active} nature of the quantum measurement process~(in contrast to the \emph{passive}---or sight-like---nature of classical measurement).  \end{inlinelist} This confluence thwarts particle reidentification during interactions, which in turn deprives of empirical cover the assumption that particles persist during interaction.  This motivates the shift of ontological focus from particles to detection events.  

This shift opens up the space to contemplate two distinct object-models of the same event-data, which provide sufficient flexibility to accommodate both approximate reidentification and the richer holistic behavior.   The doctrine of potential parts offers a provocative lens through which to view this model duality.  Moreover, this duality leads to striking consequences, such as restricted transtemporal duality, which thereby provides Schroedinger's notion of `restricted individuality'~\cite[\S11]{Schroedinger1950} and Bigaj's weak transtemporal identity~\cite[\S5]{Bigaj2020b} with a clear conceptual foundation. 

We conclude with a few open questions.

First, the conception advanced herein privileges microphysical events over microscopic objects.  But, unlike the event ontologies of, say, Russell~\cite{Russell1927} and Whitehead~\cite{Whitehead1919}, it does not extend that privilege into the macroscopic realm, since persistent and reidentifiable macroscopic objects~(rods and clocks) are operationally necessary for the space-time coordination of microphysical events\footnote{These operational considerations echo Strawson's arguments that reidentifiable objects are a precondition for linguistic reference~\cite[I.1.2]{Strawson1990}.}.  But should one regard this macro/micro distinction as fundamental or merely a consequence of the contingent fact that we need a framework of persistent objects in order to learn about microphysical reality?  

Second, the holistic object posited by the nonpersistence model confers a new kind of holism to a system of identical particles.  How does that holism relate to, or differ from, that which is often attributed to systems of entangled nonidentical particles~(e.g.~\cite{Schaffer2010,IsmaelSchaffer2020,Nager2021})?   Perhaps identical particle holism is an instance of object holism, while entanglement of nonidentical particles is~(as suggested in~\cite{Nager2021}) an instance of property holism.  If so, how do these two types of holism combine in the context of a system of entangled identical particles?

\newpage

\bibliographystyle{plainnat}

\bibliography{references_master}


\clearpage
\begin{widetext}
\begin{appendix}
\section{Reconstructed quantum symmetrization procedure~(QSP), \\ and its minimal physical interpretation}
\label{sec:rQSP-and-interpretation}

In order to facilitate comparison with the QSA given in Sec.~\ref{sec:QSA}, we here outline the reconstructed fQSA in state form, together with its minimal physical interpretation that follows directly from the reconstruction~\cite{Goyal2015} and its partial interpretation given in~\cite{Goyal2019a}.

As described in~\cite{Goyal2015}, the Feynman symmetrization postulate can be re-expressed in terms of quantum states~(see Table~\ref{tbl:QSP}).  For two structureless particles moving in one dimension,
\begin{equation} \label{eqn:SP-state}
\psiid(x_1, x_2) = \psi(x_1, x_2) \pm \psi(x_2, x_1) \quad\quad x_1 \leq x_2.
\end{equation}
Here,~$\psiid$ is the state of the system as described within the nonpersistence model, and is defined~(and normalized) over reduced configuration space,~$x_1 \leq x_2$.  The function~$\psi(x_1, x_2)$ describes the system in the persistence model and is defined over full configuration space,~$(x_1,x_2) \in \numberfield{R}^2$.

Formal extension of~$\psiid$ into the full configuration space yields
\begin{equation} \label{eqn:SP-state'}
\tpsiid(x_1, x_2) =  \frac{1}{\sqrt{2}} \bigl[ \psi(x_1, x_2) \pm \psi(x_2, x_1) \bigr],
\end{equation}
where now~$(x_1, x_2)$ ranges over~$\numberfield{R}^2$.  Formally, this equation is the same as the symmetrization postulate, but the object-duality interpretation gives it a new meaning:~the symmetrization postulate is not \emph{selection rule} that picks out allowed states~(symmetric or antisymmetric states) within a single object-model, but rather as a \emph{bridging relation} between two object-models.   Although this state is permutation invariant, this is a \emph{mathematical artefact} of the formal extension of~$\psiid$ to full configuration space, \emph{not}~(as widely asserted) a fundamental physical symmetry.

The meaning of the indices in the SP is obscured and muddled in Eq.~\eqref{eqn:SP-state'}, but can be seen clearly in Eq.~\eqref{eqn:SP-state}:~on the right-hand side of Eq.~\eqref{eqn:SP-state}, they refer to individual particles; but, on the left-hand side, they refer to \emph{locations} of detection-events.   Hence, in step~1 of the QSA, where one models a system in the persistence model~(ignoring particle identicality), the indices refer to individual particles; but in step~2, their meaning becomes model-dependent.

As mentioned in Sec.~\ref{sec:persistence-nonpersistence}, the reconstruction incorporates an isolation condition.   In Eq.~\eqref{eqn:SP-state}, when the isolation condition is satisfied,~$\psi(x_2, x_1)$ can be neglected in comparison to~$\psi(x_1, x_2)$, so that~$\psiid(x_1, x_2)$ reduces to~$\psi(x_1, x_2)$.  Hence, there is no need to impose a non-overlap~(NOR) from the outside.  When the isolation condition is satisfied, the persistence model itself suffices---thus, the particle notion becomes applicable, and the indices refer to particles.

Finally, in view of this re-interpretation, the QSA's restriction to symmetric measurement operators~(SOC) is unnecessary.  For example, the operator~$x_1$ applied to~$\psiid(x_1, x_2)$ measures the location of the left-most detection:
\begin{equation}
\ev{x_1}\rcs = \intrcsonedim x_1  \,|\psiid(x_1, x_2)|^2 dx_1\,dx_2.
\end{equation}
Similarly, the operator~$(x_2 - x_1)$ yields the distance between the detections.  When the isolation condition is satisfied, one can interpret the indices~$i = 1, 2$ as \emph{particle} labels, in which case~$x_1$ measures the location of the leftmost particle, and~$(x_2 - x_1)$ the distance between them.

%
\linespread{1.218} 
\setlength{\tabcolsep}{2pt}
\begingroup
\begin{table*}
\centering
	\begin{ruledtabular}
    \begin{tabular}{p{2.75cm}p{5.25cm}p{6.50cm}} 
    \smallskip
    {\bf }  														& \tableheadtext{Description} 			 & \tableheadtext{Remarks}    \\ \hline

\smallskip\tableheadtext{I. Persistence Model}
																	&	\smallskip\tabletext{Construct quantum persistence model of system, 
																									  treating the identical particles as individually persistent entities.}   	
																	&	\smallskip\tabletext{1. At this stage, the indices---for example in~$\psi(x_1, x_2)$---are interpreted as particle labels.}
																		\newline 
																		\tabletext{2. The classical model and the quantum persistence model share a persistent particle ontology.}
																	\\
    
\smallskip\tableheadtext{II. Synthetic Model}  
																	&	\smallskip\tabletext{Construct the synthetic quantum model of the system, which 	synthesizes the persistence and nonpersistence models of the system.}
																	&	\smallskip\tabletext{1. In the two particle example, the synthetic model yields~$\psiid(x_1, x_2) = \psi(x_1, x_2) \pm \psi(x_2, x_1)$ where~$x_1 \leq x_2$.} 
																		\newline
																		\smallskip\tabletext{2. In~$\psiid(x_1, x_2)$, the $x_i$ refer to the flash-locations~($x_1$ to the leftmost flash, $x_2$ to the rightmost), not particle locations.}
																		\newline
																		\smallskip\tabletext{3. Non-symmetric measurement operators are admissible.  For example, operator~$\hat{x}_1$ represents measurement of the location of the left-most flash.}
																		\newline
																		\smallskip\tabletext{4. If the isolation condition is valid for the experiment, then the synthetic model automatically reduces to the persistence model, and~$x_i$ can be interpreted as the location of particle~$i$.}
																	\\

	\\
    \end{tabular}
	\end{ruledtabular}
\caption{\label{tbl:QSP} \emph{Reconstructed Quantum Symmetrization Postulate~(QSP).}}
\end{table*}
\endgroup
\linespread{1.618} 

\end{appendix}
\clearpage
\end{widetext}

\end{document}